\documentclass[referee,twocolumn]{raa}           
\usepackage{graphicx,times}
\usepackage{natbib}
\usepackage{amssymb,amsmath}
\usepackage[version=4]{mhchem}
\usepackage{listings}
\bibpunct{(}{)}{;}{a}{}{,}
\usepackage{color}
\definecolor{dkgreen}{rgb}{0,0.6,0}
\definecolor{gray}{rgb}{0.5,0.5,0.5}
\definecolor{mauve}{rgb}{0.58,0,0.82}
\usepackage[a4paper=true,dvipdfm=true,pagebackref=true]{hyperref}
\hypersetup{pdftitle = The title of my PDF, pdfauthor = My name, pdfsubject= The subject, pdfkeywords = keyword1 keyword2 keyword3} 
\hypersetup{colorlinks = true, linkcolor = green, anchorcolor = red, citecolor = blue, filecolor = red, pagecolor = red, urlcolor = red}

\begin{document}
\title{GGCHEMPY: A pure Python-based gas-grain chemical code for efficient simulation of interstellar chemistry$^*$
\footnotetext{\small $*$ Supported by CASSACA.}
}

\volnopage{ {\bf 20XX} Vol.\ {\bf X} No. {\bf XX}, 000--000}
\setcounter{page}{1}

\author{Jixing Ge\inst{1}}

   \institute{Chinese Academy of Sciences South America Center for Astronomy, National Astronomical Observatories, CAS, 
Beijing 100101, People’s Republic of China. {\it gejixing666@163.com }\\
	% \and
    %	  yyy\\
    % \and 
    %     zzz\\
\vs \no
   {\small Received 20XX Month Day; accepted 20XX Month Day}
}

\abstract{In this paper, we present a new gas-grain chemical code for interstellar clouds written in pure Python (GGCHEMPY\footnote{GGCHEMPY is available on \url{https://github.com/JixingGE/GGCHEMPY}}). By combining with the high-performance Python compiler Numba, GGCHEMPY is as efficient as the Fortran-based version. With the Python features, flexible computational workflows and extensions become possible. As a showcase, GGCHEMPY is applied to study the general effects of three-dimensional projection on molecular distributions using a two-core system which can be easily extended for more complex cases. By comparing the molecular distribution differences between two overlapping cores and two merging cores, we summarized the typical chemical differences such as, \ce{N2H+}, \ce{HC3N}, \ce{C2S}, \ce{H2CO}, \ce{HCN} and \ce{C2H}, which can be used to interpret 3-D structures in molecular clouds.
\keywords{Astrochemistry --- (ISM:) evolution --- ISM: molecules --- ISM: abundances --- methods: numerical
}
}

   \authorrunning{J.X. Ge}            %author_head in even pages
   \titlerunning{The GGCHEMPY code}  % title_head in odd pages
   \maketitle

%________________________________________________ sections below
% 
\section{Introduction}           %% first-level sections will be auto-capitalized
\label{sect:intro}
The observations of molecular lines are useful tools to study the physical and chemical processes of star-forming regions such as, chemistry in the cold stage of dark clouds ($\sim 10$\,K), warm-chemistry in the warm-up stage of hot molecular cores ($\sim 10-300$\,K), and shock-derived and high-temperature chemistry in molecular outflows (a few 1000\,K) (e.g. \citealt{Herbst+2009,Bally+2016,Jorgensen+2020}). To understand the chemical processes, several astro-chemical codes using rate equation method have been developed for pure gas-phase reactions (e.g. \citealt{McElroy+etal+2013+UMIST,Wakelam+2014+Nahoon}), gas-phase reactions and accretion/desorption processes related to dust grains (e.g \citealt{Maret+Bergin+2015+astrochem}), full gas-grain processes with dust surface reactions (e.g. \citealt{Hasegawa+etal+1992+two,Hasegawa+1993+three,Garrod+2006,Garrod+2008, Semenov+etal+2010,Grassi+etal+2014+KROME,Ge+etal+2016b,Ge+etal+2016a,Ge+etal+2020b,Ge+etal+2020a,Ruaud+etal+2016,Holdship+etal+2017+UCLCHEM}).

Generally, the astrochemical model is a single-point one (0-D) with fixed physical parameters and a chemical reaction network. It can be applied to 1-D, 2-D and 3-D cases by sampling points with corresponding physical parameters of a complex physical structure which may need multiple steps to prepare a model for a specific case. Thus, the simplicity of the use of the code becomes an important point. Another point is that the codes written in Fortran/C are dependent on the compiler and the libraries installed in the computer system, such as the ordinary differential equation (ODE) solvers: ODEPACK\footnote{\url{https://people.sc.fsu.edu/~jburkardt/f77_src/odepack/odepack.html}} in Fortran or CVODE\footnote{\url{https://computing.llnl.gov/projects/sundials/cvode}} in C. For example, some Fortran codes developed with gfortran may report errors when compiled by the Intel Fortran compiler (ifort) due to some minor syntax differences. Moreover, users should pre-build their computer environments which vary among different operating systems (Linux/Windows/Mac OS). This may be not good for some astronomers because that $\sim 63\pm 4$\% astronomers have not taken any computer-science courses at an undergraduate or graduate level as reported by a survey of astronomers (\citealt{Bobra+etal+2020}). Therefore, a simple but powerful astro-chemistry code is desired. 

With increasing Python users in astronomy (see e.g. $\sim 67\pm 2$\% reported by \citealt{Momcheva+2015} and $\sim 66$\% by \citealt{Bobra+etal+2020}), the Python programming language is likely a good choice to solve the above issues due to its flexible syntax and power package organization, which makes Python to be system-independent. In this work, we present a gas-grain chemical code written in pure Python (GGCHEMPY) to fetch the flexible features of Python and to reach comparable speed to the existing Fortran version by using Python package Numba\footnote{\url{http://numba.pydata.org/}.} which translates Python functions to optimized machine code at runtime using the industry-standard low level virtual machine (LLVM) compiler library. As a showcase, GGCHEMPY is applied to discuss the three-dimensional projection effects of molecular cloud cores on molecular distributions using a two-core system with typical physical conditions of Planck galactic cold clumps (PGCCs). This serves as a complementary work of \cite{Ge+etal+2020a} which discusses the molecular projection effects in a complex and specific PGCC G224.4-0.6. 

This paper is organized as follows. Section~\ref{sec:code} describes the gas-grain code GGCHEMPY. In Section~\ref{sec:showcase}, the GGCHEMPY code is applied to discuss a more general projection effect on molecular distributions. Finally, a summary is given in Section~\ref{sec:summary}.

\section{Chemical model and code} \label{sec:code}
In this section, we describe the Gas-Grain CHEMistry code written in pure Python (GGCHEMPY). It was developed on basis of the gas-grain chemical processes described in literature (e.g.  \citealt{Hasegawa+etal+1992+two,Semenov+etal+2010}) and the Fortran\,90 version of GGCHEM that have been used for various chemical problems in interstellar clouds (e.g \citealt{Ge+etal+2016b,Ge+etal+2016a,Tang+etal+2019,Ge+etal+2020b,Ge+etal+2020a}).

\subsection{Chemical model}
In the model, the gas-grain chemical processes include gas-phase reactions and 
dust surface reactions which linked through accretion and desorption processes of neutral species (e.g. \citealt{Hasegawa+etal+1992+two,Semenov+etal+2010}). The gas-grain reaction network\footnote{The network was downloaded from KIDA database: \url{http://kida.astrophy.u-bordeaux.fr/networks.html}} from \cite{Semenov+etal+2010} is updated and used in this work. Beside the thermal and cosmic-ray-induced desorption processes (\citealt{Hasegawa+1993+CR}), the reactive desorption (\citealt{Garrod+etal+2007,Minissal+etal+2016}) 
and CO and \ce{H2} self-shielding (\citealt{Lee+etal+1996}) are implemented. 

The GGCHEMPY code simulates the gas-grain chemical processes by solving the Ordinary Differential Equations (ODEs) of a species $i$ in gas-phase:
\begin{equation}
	\frac{\mathrm{d} n_{i}}{\mathrm{d} t}=\sum_{l, m} k_{l m} n_{l} n_{m}-n_{i} \sum_{i \neq l} k_{l} n_{l}+k_{i}^{\mathrm{des}} n_{i}^{s}-k_{i}^{\mathrm{acc}} n_{i},
\end{equation}
and the corresponding neutral species on dust surface
\begin{equation}
	\frac{\mathrm{d} n_{i}^{s}}{\mathrm{d} t}=\sum_{l, m} k_{l m}^{s} n_{l}^{s} n_{m}^{s}-n_{i}^{s} \sum_{i \neq l} k_{l}^{s} n_{l}^{s}-k_{i}^{\mathrm{des}} n_{i}^{s}+k_{i}^{\mathrm{acc}} n_{i},
\end{equation}
where $n$ is the number density of species and $k$ is the reaction rate coefficient. The superscripts 's', 'des' and 'acc' indicate surface species, desorption process and accretion process respectively. The formulas to compute reaction rate coefficient can be found in paper of \cite{Semenov+etal+2010}. The Python package \texttt{scipy.integrate.ode} is used to do the integration using the backward-differentiation formulas (BDF) method.

\subsection{Usage and benchmark of GGCHEMPY} \label{sec:frame}
To avoid duplication of variable name when being combined with other codes (such as radiation transferring code), GGCHEMPY is modularized into Python classes. There are seven attributes of \texttt{ggchempylib.ggchempy}: \texttt{gas, dust, ggpars, elements, species, reactions, iswitch}. Thus, physical and chemical properties can be defined separately in the classes such as gas temperature \texttt{gas.T}, dust temperature \texttt{dust.T}, species mass \texttt{species.mass} etc. The class \texttt{ggpars} is used to store common parameters, such as constants, parameters for reaction network, parameters for ODEs, controls of time steps etc. The class \texttt{iswitch} is used to switch on some functions, such as \ce{H2} and CO self-shielding (\citealt{Lee+etal+1996}) (\texttt{iswitch.iSS=1}) and reactive desorption from \cite{Garrod+etal+2007} (\texttt{iswitch.iNTD=1}) or \cite{Minissal+etal+2016} (\texttt{iswitch.iNTD=2}). To switch off a function, just pass "0" to the class.

The GGCHEMPY code can be downloaded from \url{https://github.com/JixingGE/GGCHEMPY}. To install GGCEHMPY, type the following command on the terminal:\newline
\texttt{python setup.py build}\newline
\texttt{python setup.py install}\newline
After installation, GGCHEMPY can be activated via a graphical user interface (GUI) or Python script. To use the GUI, just type \texttt{run\_GUI()} after importing  it from GGCHEMPY (e.g. \texttt{from ggchempylib import run\_GUI}). Three necessary steps are needed to run a model in a python script:
\begin{itemize}
\item[(1)] Import necessary functions, such as \texttt{from ggchempylib import ggchempy}
\item[(2)] Set parameters according to your model in the python script
\item[(3)] Run the code by typing \texttt{ggchempy.run(modelname)}
\end{itemize}

GGCHEMPY works by calling \texttt{init\_ggchem()} to initialize reaction network according to your input parameters. Then it computes reaction rate coefficients for all reactions by calling \texttt{compute\_reaction\_rate\_coefficients()}. Finally, to solve the ODEs, BDF method from \texttt{scipy.integrate.ode} is used to do the integration. Here, the Numba decorator \texttt{@numba.jit(nopython=True)} is added to the Python functions \texttt{fode(y,t)} and \texttt{fjac(y,t)} to accelerate the integration of ODEs, which need massive calculations and loops. The modeled results will be saved into \texttt{"out/DC.dat"} if \texttt{modelname="DC"} and \texttt{ggchempy.ggpars.outdir="out/"}. The modeled results can also be directly passed to a Python dictionary (\texttt{DC}) for analysis/plot in the same Python script by typing e.g. \texttt{DC=ggchempy.run("DC")}.

\begin{figure}[ht!]
	\centering
	\includegraphics[width=0.9\textwidth]{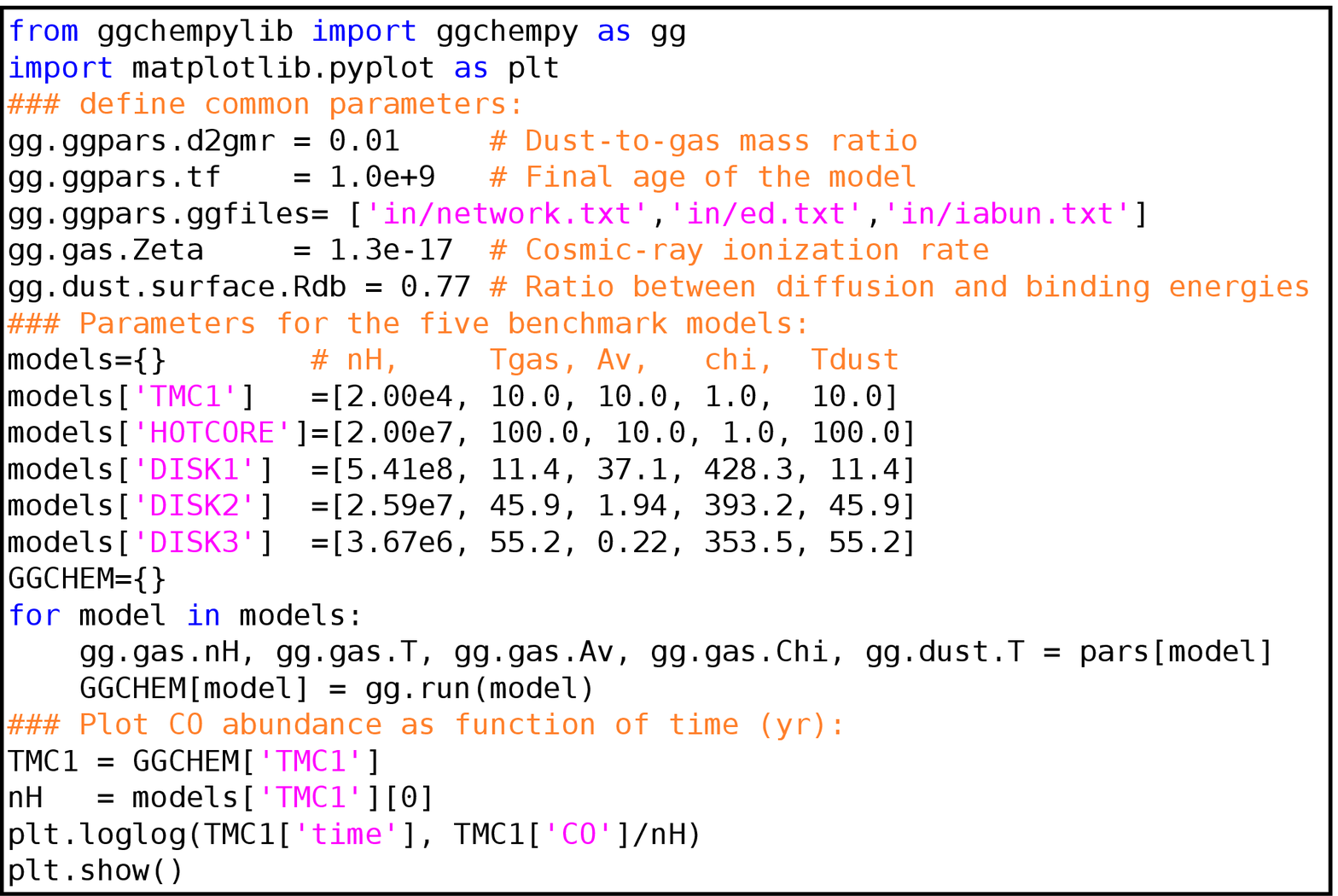}
	\caption{An exemplar python script to use GGCHEMPY to compute five benchmark chemical models and explore the results.}
	\label{fig:code}
\end{figure}
\begin{figure}[ht!]
\centering
\includegraphics[width=0.5\textwidth]{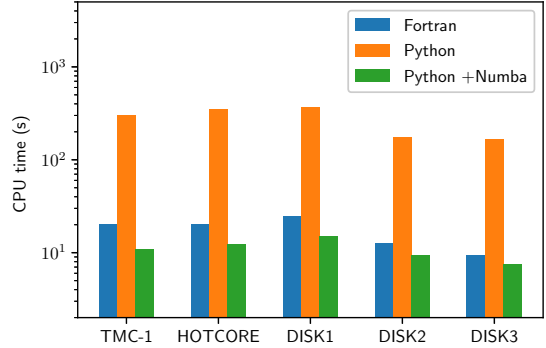}
\caption{Comparison of CPU times used by the Fortran code (GGCHEM, blue) and GGCHEMPY (without (orange) and with Numba (green)) for the five benchmark models of TMC1, HOTCORE, DISK1, DISK2 and DISK3. }
\label{fig:bench2010}
\end{figure}

The benchmark of the GGCHEMPY code was successfully made with the five models from \cite{Semenov+etal+2010} using the Python script shown in Fig.~\ref{fig:code}. A laptop with Linux system and Intel i7-3612QM (4 cores, 2.1\,GHz) was used. The Python version is 3.8.5. For one model, GGCHEMPY can finish within about 10 seconds which is comparable to the Fortran version, see Fig~\ref{fig:bench2010}. The benchmark results and screenshot of GUI and the full documents can be found on the GGCHEMPY webpage. 

\section{A showcase: three-dimensional projection effects} \label{sec:showcase}
As a showcase in this section, GGCHEMPY is applied to combine with physical models to study general three-dimensional projection effects on molecules (TDPEs). The TDPE was proposed by \cite{Ge+etal+2020a} for a Planck galactic cold clump (PGCC) G224.4-0.6 with four cores interpreted from observations which is a special one constrained by observed molecules of \ce{N2H+}, \ce{HC3N} and \ce{C2S}. The MHD simulation of filamentary molecular clouds made by \citealt{Li+2019} could also support the TDPEs. Considering the TDPEs could commonly exist in space, we explore a more general TDPE with more molecules. In addition, the chemical differences between the overlapping cores and the merging cores are not explored yet which are important for interpreting observations and are also the goals in this section.

To study general TDPEs, typical physical conditions of PGCCs are adopted because that 
\begin{itemize}
\item The PGCCs have typical conditions of $T_{\rm dust}\sim 7-20$\,K and $N_{\rm H_2}\sim 10^{20}-5\times 10^{22}$\,cm$^{-2}$ (e.g. \citealt{Planck+2011a,Planck+2011,Planck+2016,Wu+2012,Mannfors+etal+2021}). They have been continuously studied via observations by many projects (e.g. \citealt{Wu+2012,Liu+2012, Liu+2018, Tatematsu+2017, Tang+2018,Tang+etal+2019, Yi+2018, Eden+2019, Dutta+2020,Sahu+2021, Yi+2021,Wakelam+etal+2021}) with notable molecular peak offsets from continuum peaks, such as \ce{N2H+}, \ce{HC3N} and \ce{C2S} in samples of PGCCs (\citealt{Tatematsu+2017,Tatematsu+etal+2021}) and \ce{C2H} in a PGCC G168.72-15.48 (\citealt{Tang+etal+2019}). Therefore, they are ideal sources to verify the TDPEs through observations. 

\item The typical physical conditions of PGCCs provide observable information for our large JCMT project SPACE\footnote{{\bf SPACE} project information: {\bf Title}: "Submillimeter Polarization And Chemistry in Earliest star formation (SPACE). {\bf PI}: Dr. Tie Liu. {\bf Webpage}: \url{https://www.eaobservatory.org/jcmt/science/large-programs/space/}".} in which one of the scientific goals is to verify the TDPEs through observable molecular distributions. 
\end{itemize}

In the following sections, physical models are described in Section~\ref{sec:physical_models}. We describe spherical core with the Plummer-like physical structure in Section~\ref{sec:SCC} which is used as a basic unit to build the following two-core systems: two overlapping cores along line-of-sight in Section~\ref{sec:TOCC} and a face-on two merging cores in Section~\ref{sec:TMCC} as a comparison. Results and discussions are shown in Section~\ref{sec:results} and \ref{sec:discussions} respectively. Considering the complexity of 3-D structures and the purpose of the showcase, we do not explore more parameter space, such as the variation of parameters to build a Plummer-like core, the viewing angle of the two merging cores etc. 
To facilitate the comparison with the two-overlapping-cores model, we only consider the orientation of the two-merging-cores model when both cores are in the same sky plane and we call it "face-on".

\subsection{Physical models}\label{sec:physical_models}
\subsubsection{Spherical cloud core (SCC)}
\label{sec:SCC}
\begin{figure*}[ht!]
\centering
\includegraphics[scale=0.55]{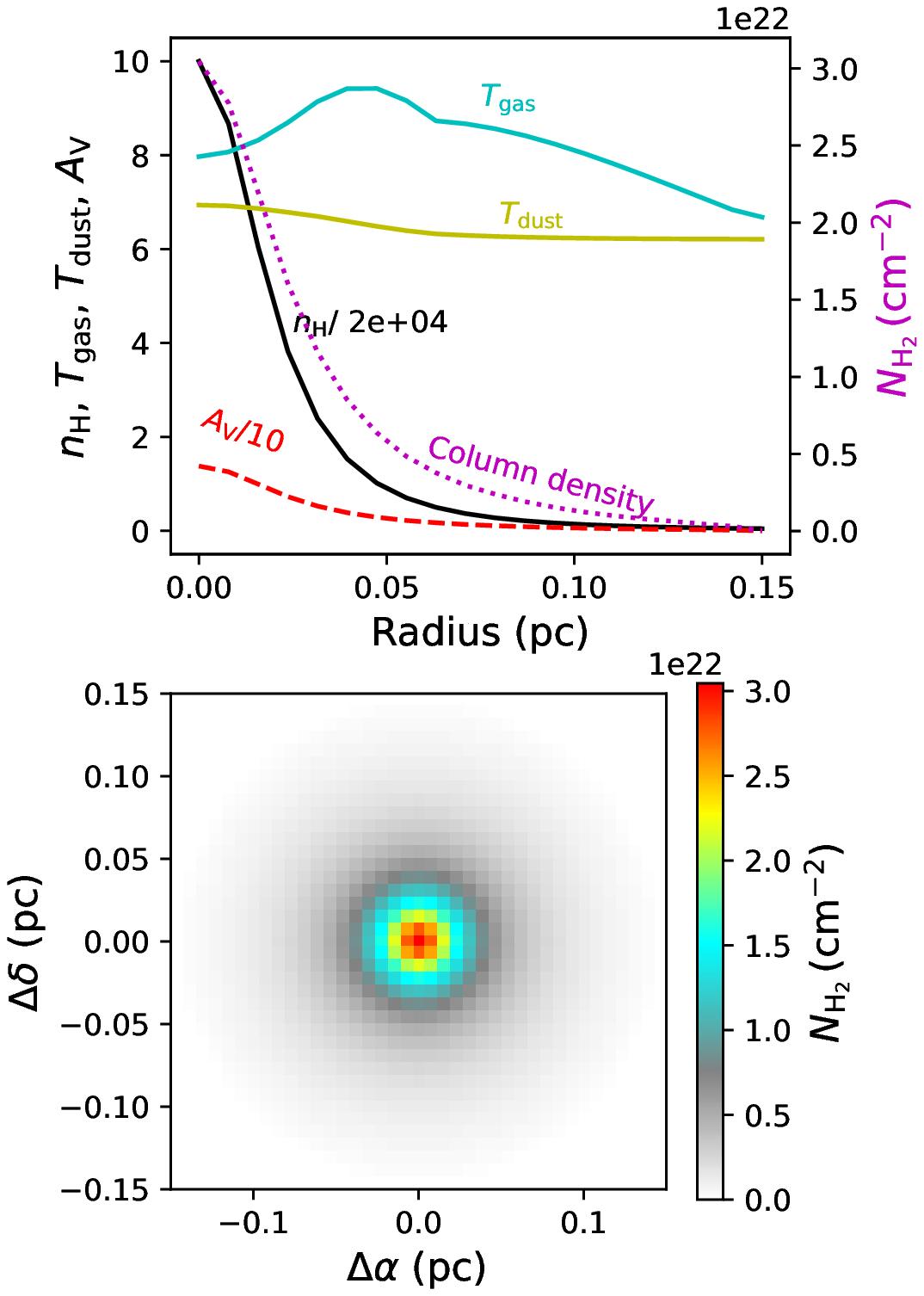}
\includegraphics[scale=0.46]{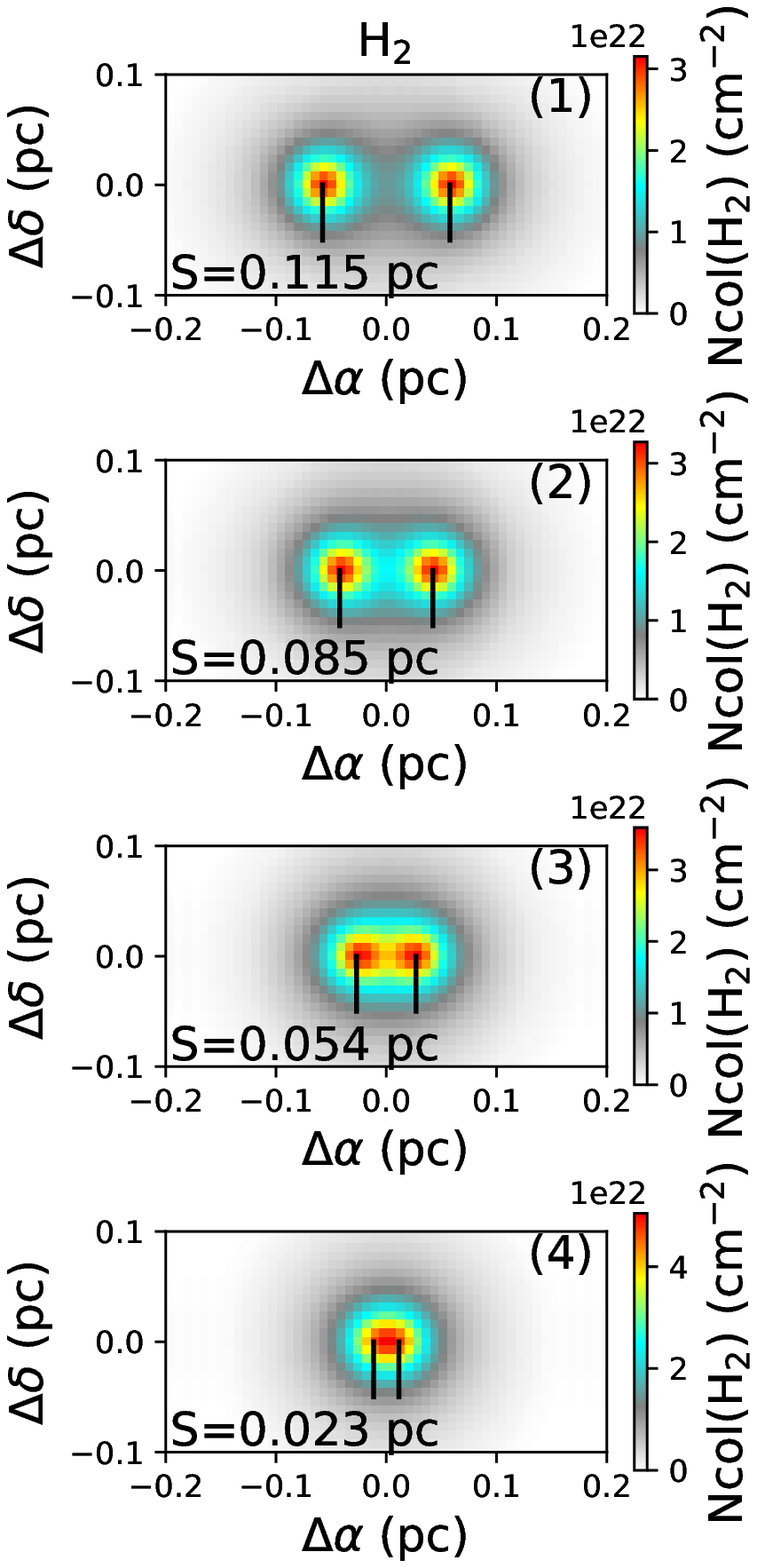}
\caption{Left-top: radial profiles of density (cm$^{-3}$), extinction (mag), gas and dust temperatures (K). Left-bottom: \ce{H2} column density map. Right:  \ce{H2} column density maps of two overlapping cores with different separations ($S$, values shown with labels and marked by vertical lines) from top to bottom for S1 to S4, which serve as continuum maps. Right panels with limited ranges of axes to show central part.}
\label{fig:radial_profiles_phy}
\end{figure*}
For a spherical cloud core, we use the 1-D radial density profile described by the Plummer-like function (\citealt{Plummer+1911})
\begin{equation}
n(r) = \frac{n_c}{ (1+(r/r_c)^2 )^{p/2}}
\label{eq:plummer}
\end{equation}
where $n_c$ is the central density, $r_c$ is the central flat region and $p$ is the pow-law index. 

We set the needed parameters by Eq.(\ref{eq:plummer}) according to the following physical conditions of PGCCs. For PGCCs in $\lambda$ Orionis cloud, Orion A and B, \cite{Yi+2018} reported the cloud cores with [$R=0.08$\,pc, $n_{\rm H}=(2.9\pm 0.4)\times 10^5$\,cm$^{-3}$], [$R=0.11$\,pc, $n_{\rm H}=(3.8\pm 0.5)\times 10^5$\,cm$^{-3}$] and [$R=0.16$\,pc, $n_{\rm H}=(15.6\pm 1.8)\times 10^5$\,cm$^{-3}$] in which $R$ is the radius of the core. For PGCCs in the L1495 dark cloud, \cite{Tang+2018} derived central densities of $n_c\sim 1.3\times 10^4-1.8\times 10^5$\,cm$^{-3}$ with $r_c\sim 0.01-0.1$ and outer radius of $R\sim 0.06-1.0$\,pc. Considering the above parameters derived from observations, we set $n_c$, $r_c$ and $R$ in our model to be $2\times 10^5$\,cm$^{-3}$, 0.025\,pc and $0.15$\,pc respectively. The $p$ value is set to be 3.0 which is an intermediate one (e.g. $p\sim 1.5-4.3$ for the four PGCCs used in the work of \cite{Ge+etal+2020a}). 

We sample 20 radial points with a resolution of $1628$\,AU for the 1-D radial physical profile. Thus, a cube with $39\times 39\times 39$ points is built for a spherical cloud core by interpolating on the 1-D physical profile. The 2-D column density map $N_{\rm H_2}(x,y)$ is obtained by integrating along the line of the sight ($z$). In Fig.~\ref{fig:radial_profiles_phy}, the left-top and left-bottom panels show the radial physical profiles and the \ce{H2} column density map respectively, showing that the values at the center and edge are about $3.06\times 10^{22}$ and $2.95\times 10^{20}$\,cm$^{-2}$ respectively. 

The visual extinction is estimated using the relation to \ce{H2} column density $N_{\rm H_2}(r)$ (\citealt{Guver+2009}) via 
\begin{equation}
A_{\rm V} = N_{\rm H_2}(r)/2.21\times 10^{21},
\label{eq:Av}
\end{equation}
Here we neglect the mutual shielding effects between the two cores. The extinctions vary from $\sim 13.80$\,mag (center) to $\sim 0.13$\,mag (edge). The radial gas and dust temperatures are estimated using the formulas from \cite{Goldsmith+2001}, considering cooling and heating processes of molecules and the effects of coupling between the gas and the grains. The extinction and temperature profiles are shown in the left-top panel of Fig.~\ref{fig:radial_profiles_phy}. 

\subsubsection{Two overlapping cloud cores (TOCC)}\label{sec:TOCC}
\begin{figure}[ht!]
\centering
\includegraphics[width=0.45\textwidth]{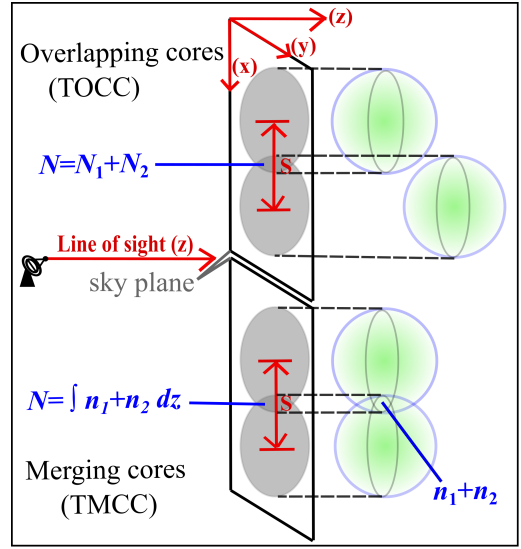}
\caption{Sketches of the two overlapping cloud core model (TOCC, upper part) and the two face-on merging cloud core model (TMCC, lower part). $n$ and $N$ are number density and column density respectively.}
\label{fig:models}
\end{figure}

To study the TDPEs, we adopt a simple model with two same spherical cores with an adjustable separation ($S$) on the sky plane (see the upper part of Fig.~\ref{fig:models}). 
The \ce{H2} column density map of the two overlapping cloud cores (hereafter TOCC) is reached by summing the column density maps of the two cores with an assumption of optically thin and an adjustable separation ($S$): $N_{\rm H_2} = \int_{-R}^{+R} n_1(x,y,z) dz + \int_{-R}^{+R} n_2(x,y,z) dz$, where $z$ is the line-of-sight direction. Four $S$ values are adopted and shown with labels together with the \ce{H2} column density maps in the right panels of Fig.~\ref{fig:radial_profiles_phy}. The modeled peak values of \ce{H2} column density is about $3-5\times 10^{22}$\,cm$^{-2}$ which is consistent with the lower limits ($2.5\times 10^{22}$\,cm$^{-2}$) derived from a sample of PGCCs (\citealt{Yi+2018}).

\subsubsection{Two merging cloud cores (TMCC)}\label{sec:TMCC}
We also build a model with two merging cloud cores at the same sky-plane to compare the chemical differences with the TOCC model (see the lower part of Fig.~\ref{fig:models}). Since the gravitational stability of such a system is not the concern of this work, the physical models will stay unchanged during the chemical model evolution. At the merged part, the gas densities are first summed before running chemical models which results in a \ce{H2} column density map with $N_{\rm H_2} = \int_{-R}^{+R} (n_1(x,y,z) + n_2(x,y,z)) dz$. For estimating the visual extinction, we simply calculate six values from six directions at each grid point and adopt the minimum one but timing a factor of 2.0 to match the one derived from Eq.(\ref{eq:Av})) along the line of sight.

\subsection{Chemical structure}
To simulate the chemical structures of the above physical models, GGCHEMPY is called as a Python module. For an SCC model, chemical models with 20 radial points can be finished within CPU time of about $200=20\times 10$\,seconds. For the TOCC model, two same chemical structures of the SCC model can be used with an adjustable separation under the optically thin assumption. However, for the TMCC model, a large number of single-point chemical models are needed to build a 3-D chemical structure because that the densities should be first summed at the merged parts. For example, about 14 days is needed for a 3-D grid of $39\times 78 \times 39$ with 10 seconds for one model. Even considering the symmetry, the CPU time can only be reduced from 14 days to ~1.7 days (14/8). In addition, we also need to change the merged parts according to the varying separation which needs more time. To reduce the computation time, we build a mode grid with varied density, extinction and temperature which results in about 700 models and needs about 3 hours to run (see details in Appendix~\ref{app:grid}). Once the models on the grids are finished, only $\sim 15$ seconds are needed to load them. Thus, the modeled abundances at any given parameters can be obtained by interpolating on the grid. The interpolated values are compared with the true model with explicit parameters in Fig.~\ref{fig:grid:comparison} in Appendix~\ref{app:grid}. In this way, the computation time is highly reduced. For the TMCC models, we use a fixed temperature of 10\,K for both gas and dust grains. Comparing to the models with varied gas and dust temperatures, we find that the temperature effects on molecular abundances are very small, see discussions in Section~\ref{sec:temperature}. 

\subsection{Results}\label{sec:results}
Typical molecules are discussed such as, \ce{N2H+}, \ce{HCO+}, \ce{HCN}, \ce{HC3N}, \ce{H2CO}, \ce{C2H}, \ce{C2S} and SO, in which some of them are tracers of the SPACE project. To simulate molecular column density maps, we turn on the switch of GGCHEMPY to include the reactive desorption (hereafter RD) proposed by \cite{Minissal+etal+2016} which is important for the cloud core models. The column density maps of molecules are obtained by combining the chemical models on the TMCC and TOCC physical model grids at given ages. Fig.~\ref{fig:N2H+}-\ref{fig:C2H} show the modeled column density maps of selected species. In these figures, the grayscale shows the \ce{H2} column density map. The cyan and red contours show the maps from TMCC and TOCC models respectively with levels of $(0.3\,, 0.6\,, 0.9)\times N_{\rm max}$ where $N_{\rm max}$ is the maximum column density of a species. Comparing the modeled peak column densities at age of $5\times 10^5$\,yr with the observed ones of \ce{N2H+} ($\sim 10^{12}-10^{13}$), \ce{HCN} ($\sim 10^{12}-10^{14}$), \ce{H2CO} ($\sim 10^{13}-10^{14}$) and \ce{C2H} ($\sim 10^{14}-10^{15}$) in PGCCs within $\lambda$ Orionis, Orion A and B clouds (\citealt{Yi+2021}), we found that good agreements are reached within typically one order of magnitude. The panel ID in form of (Sn,age) is to indicate the modeled result with a given separation from the four sets (n=1,2,3,4) defined in Section~\ref{sec:TOCC} and a given age.

\subsubsection{Chemical differences between TMCC and TOCC models}
\begin{figure}[ht!]
\centering
\includegraphics[width=0.3\textwidth]{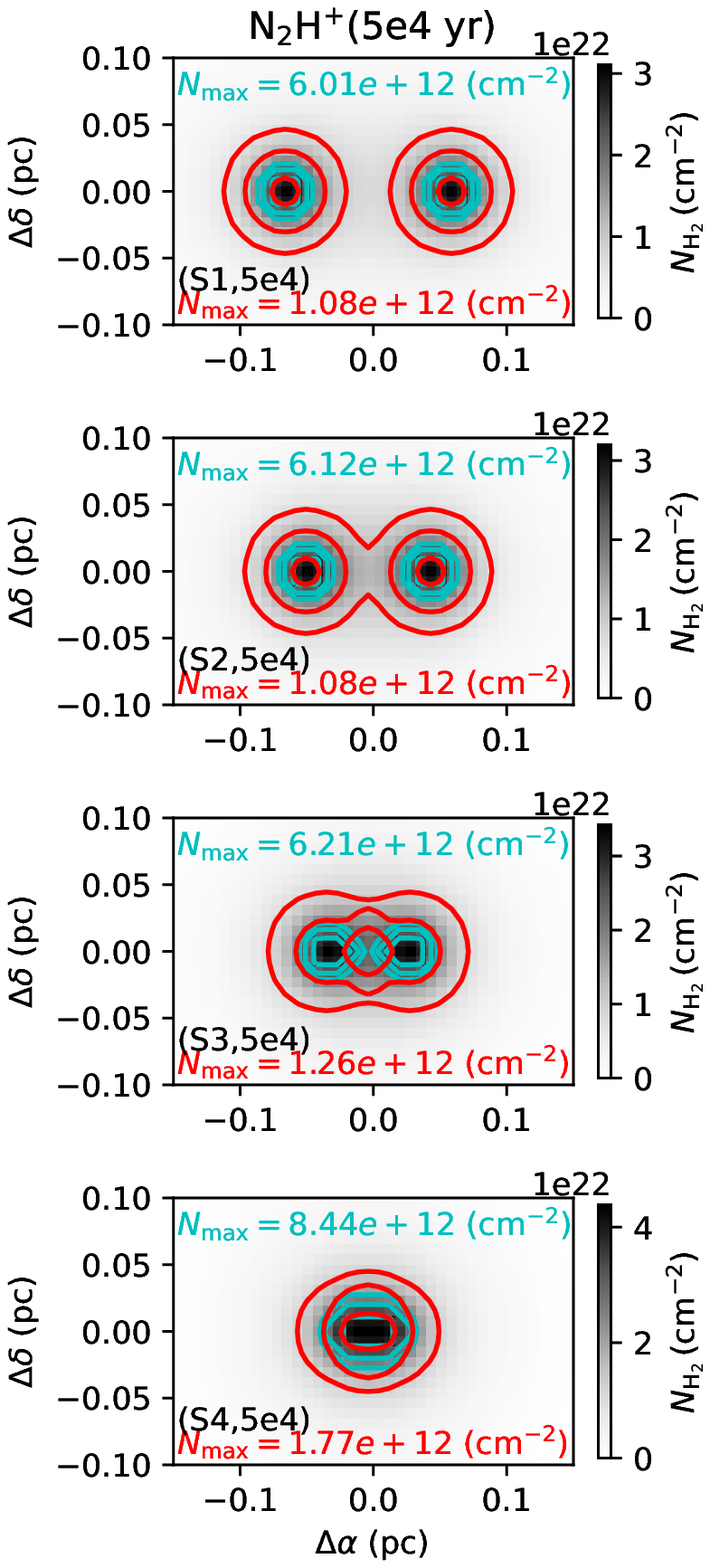}
\includegraphics[width=0.3\textwidth]{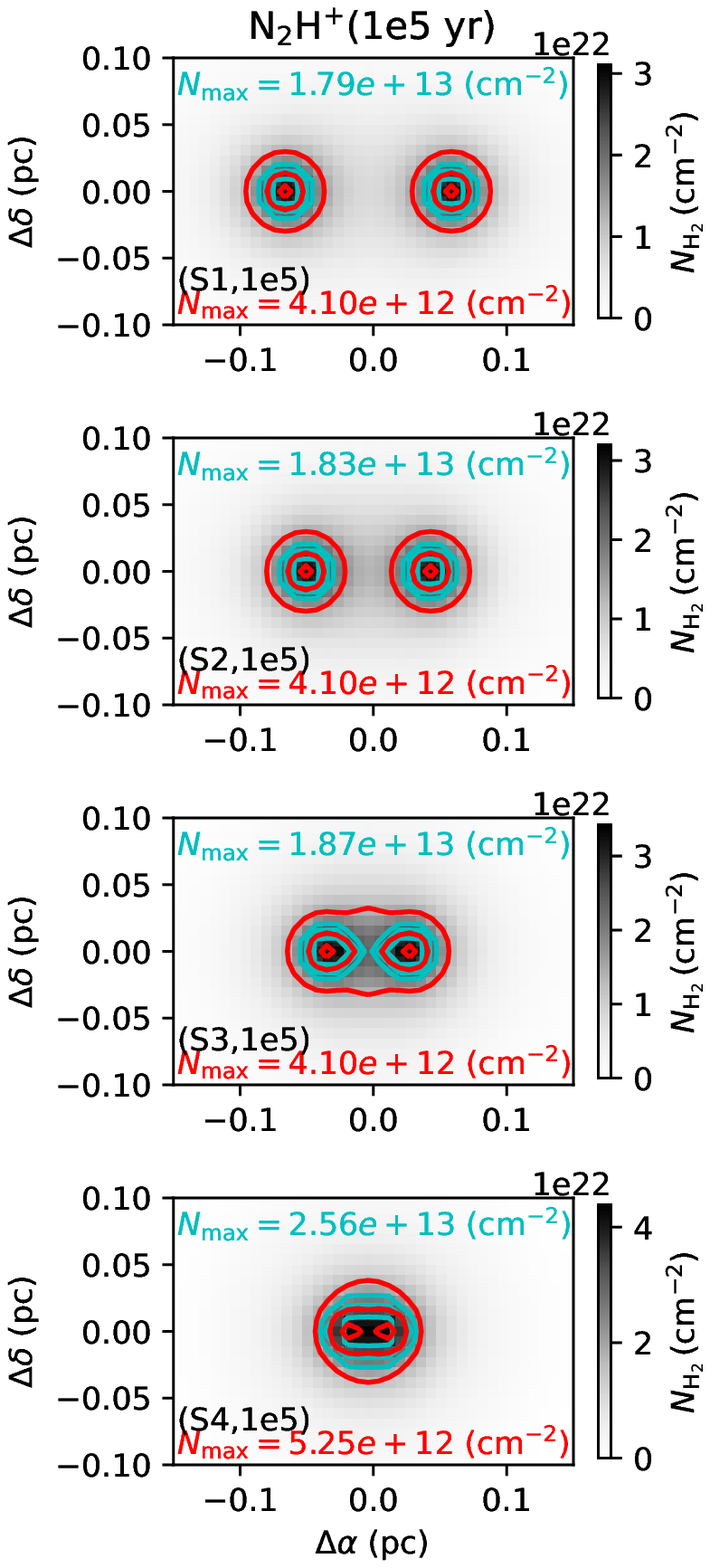}
\includegraphics[width=0.3\textwidth]{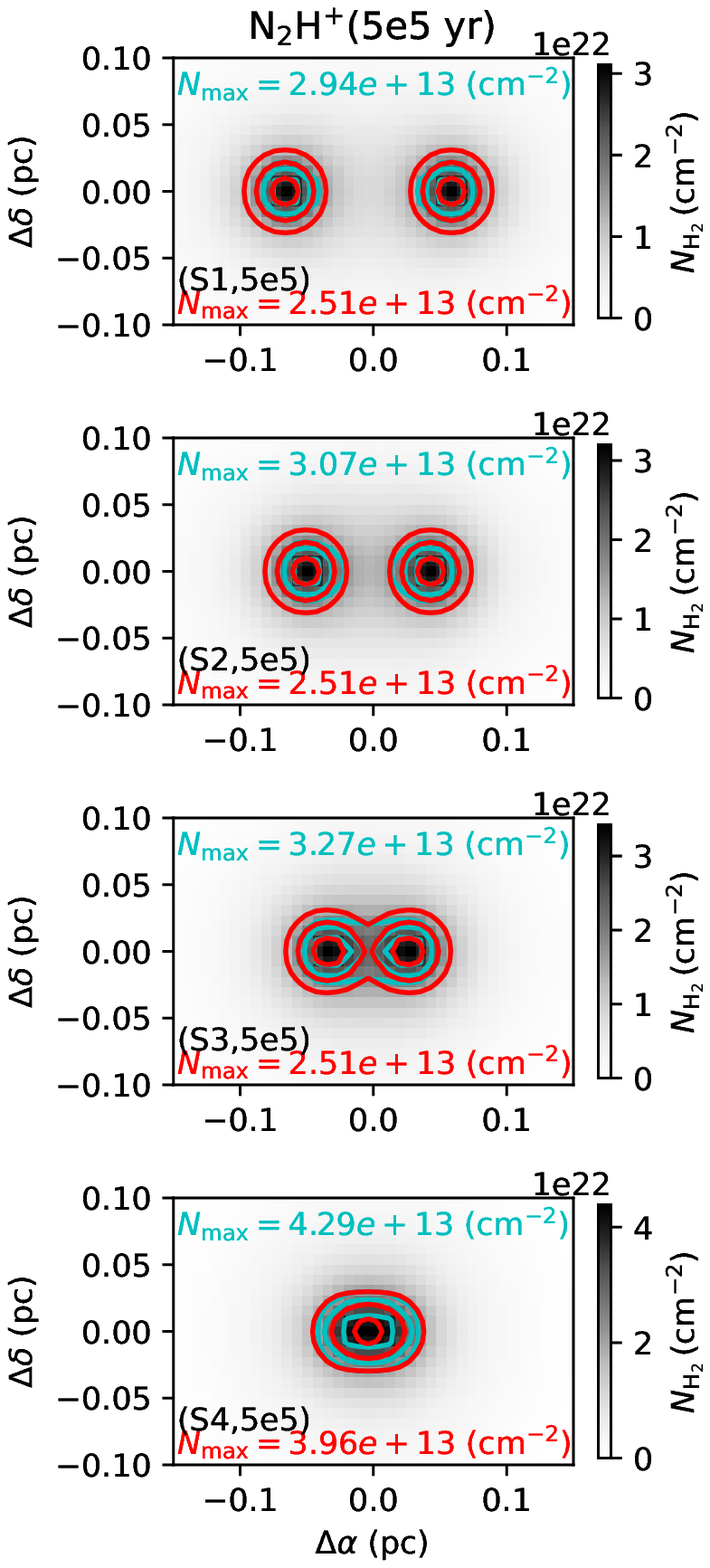}
\caption{Modeled \ce{N2H+} column density maps from TMCC model (cyan contours) and TOCC model (red contours) overlapping on \ce{H2} column density map (gray scale), with varied ages of $5\times 10^4$, $10^5$, $5\times 10^5$\,yrs for left, middle and right panels respectively. The contours with levels of $(0.3\,, 0.6\,, 0.9)\times N_{\rm max}$ where $N_{\rm max}$ is the maximum column density of \ce{N2H+} shown as texts. }
\label{fig:N2H+}
\end{figure}

For \ce{N2H+} in Fig.~\ref{fig:N2H+}, the TOCC model and the TMCC model have similar distributions except for the one in panel (S3,5e4). This is because that the depletion effects of \ce{N2H+} is very small due to its parent species \ce{N2} having high gas-phase abundance (\citealt{Womack+etal+1992}). The \ce{N2H+} peak in the TOCC model (red) in panel of (S3,5e4) is due to superposition of the flat abundance distributions of \ce{N2H+} in the sky area between the two cores. This occurs when the size of the \ce{N2H+} central flat region is comparable to the separation between the two cores.

\begin{figure}[ht!]
\centering
\includegraphics[width=0.3\textwidth]{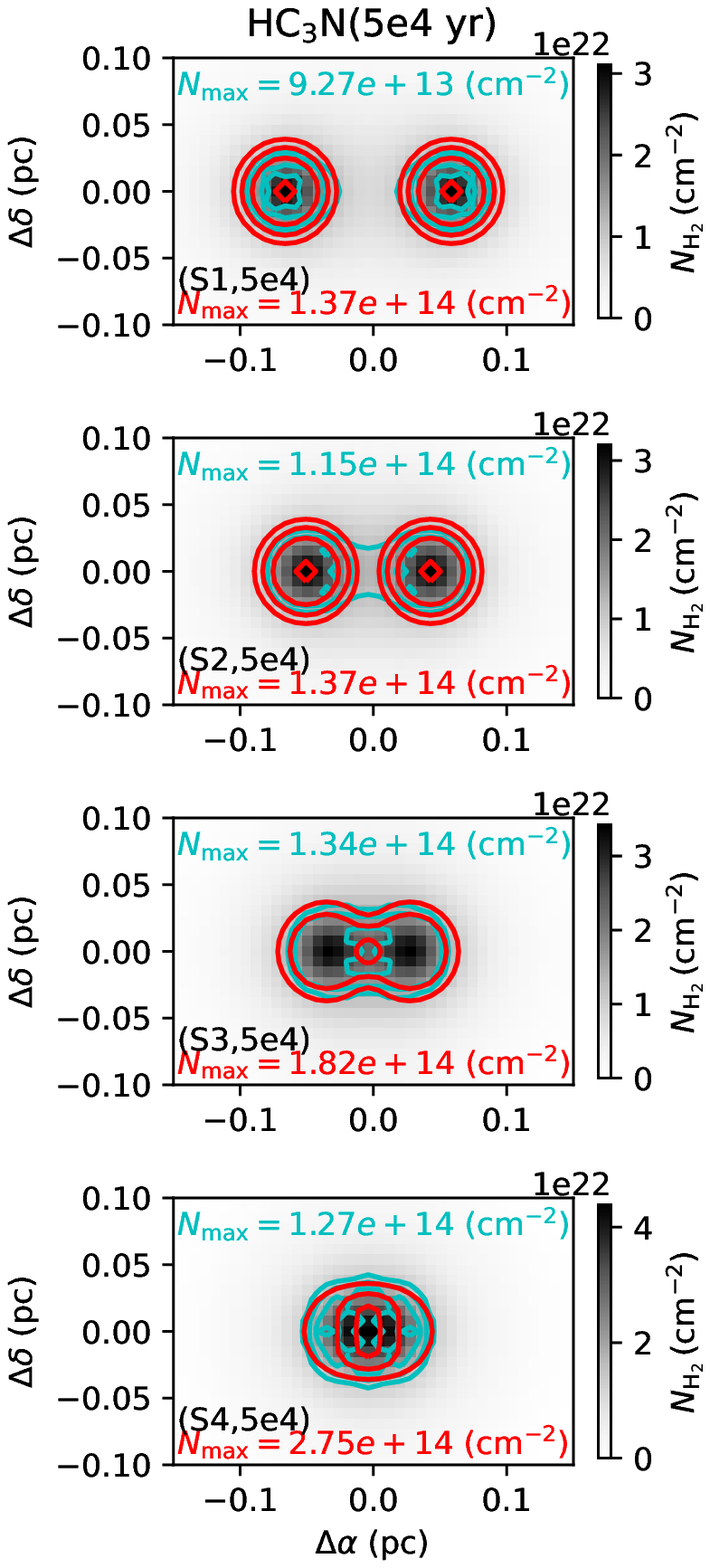}
\includegraphics[width=0.3\textwidth]{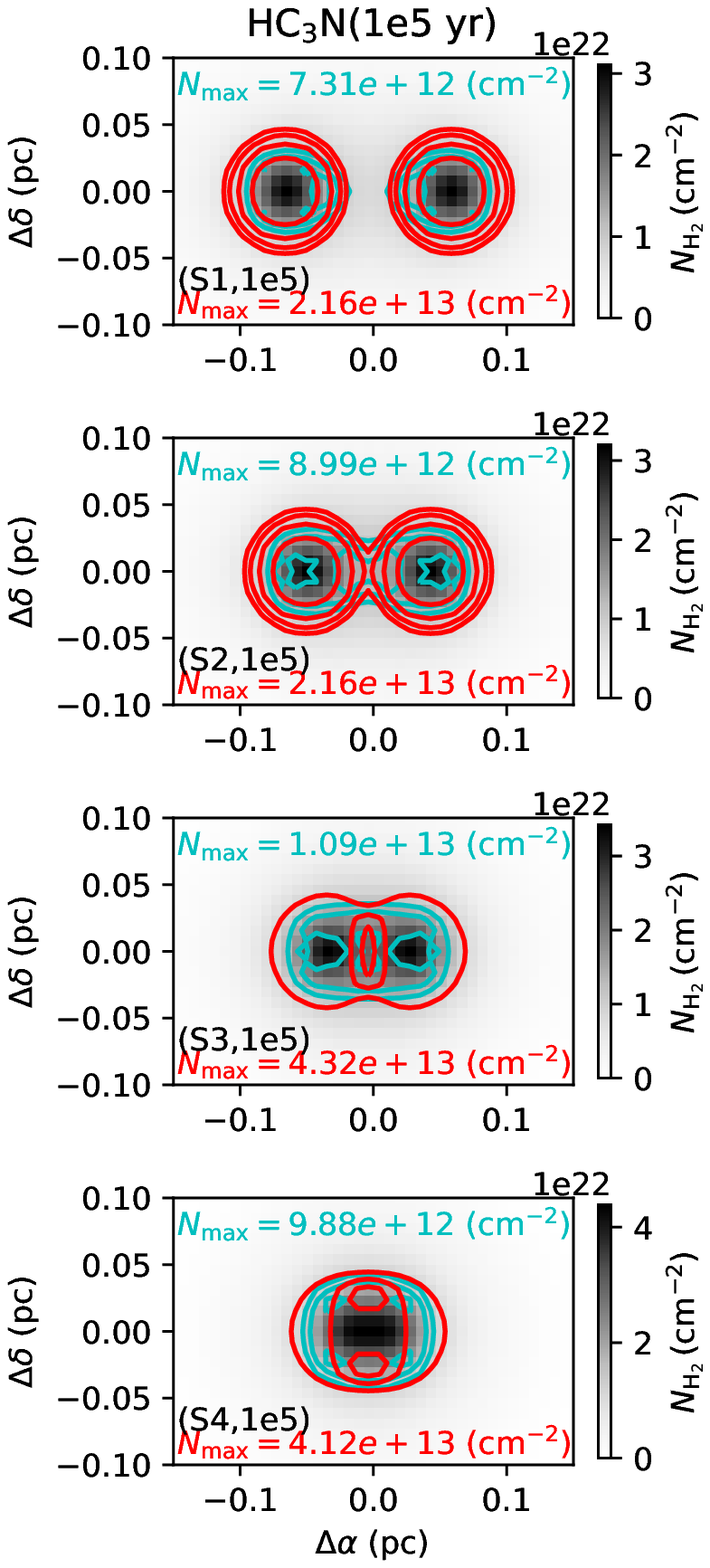}
\includegraphics[width=0.3\textwidth]{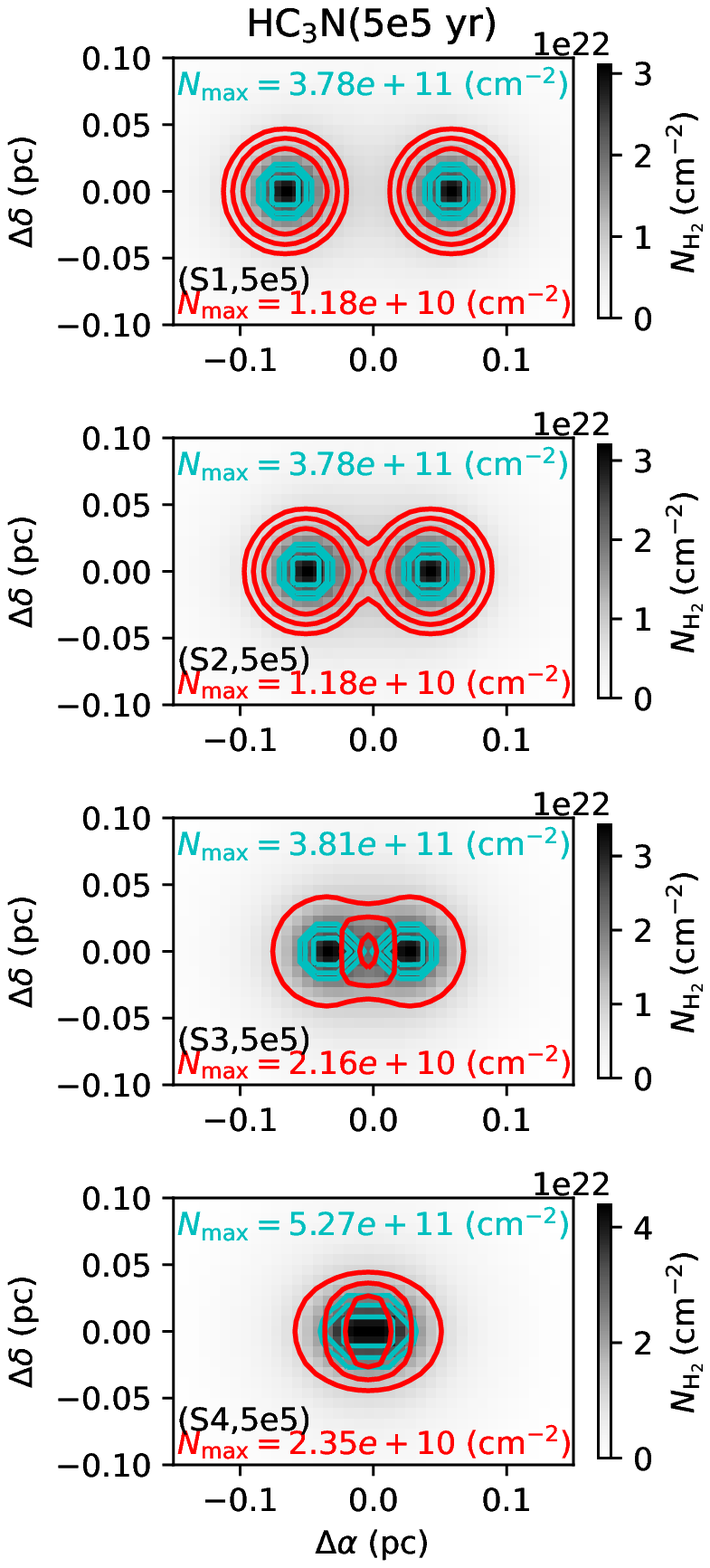}
\caption{Same as Fig.~\ref{fig:N2H+}, but for \ce{HC3N}.}
\label{fig:HC3N}
\end{figure}

\begin{figure}[ht!]
\centering
\includegraphics[width=0.3\textwidth]{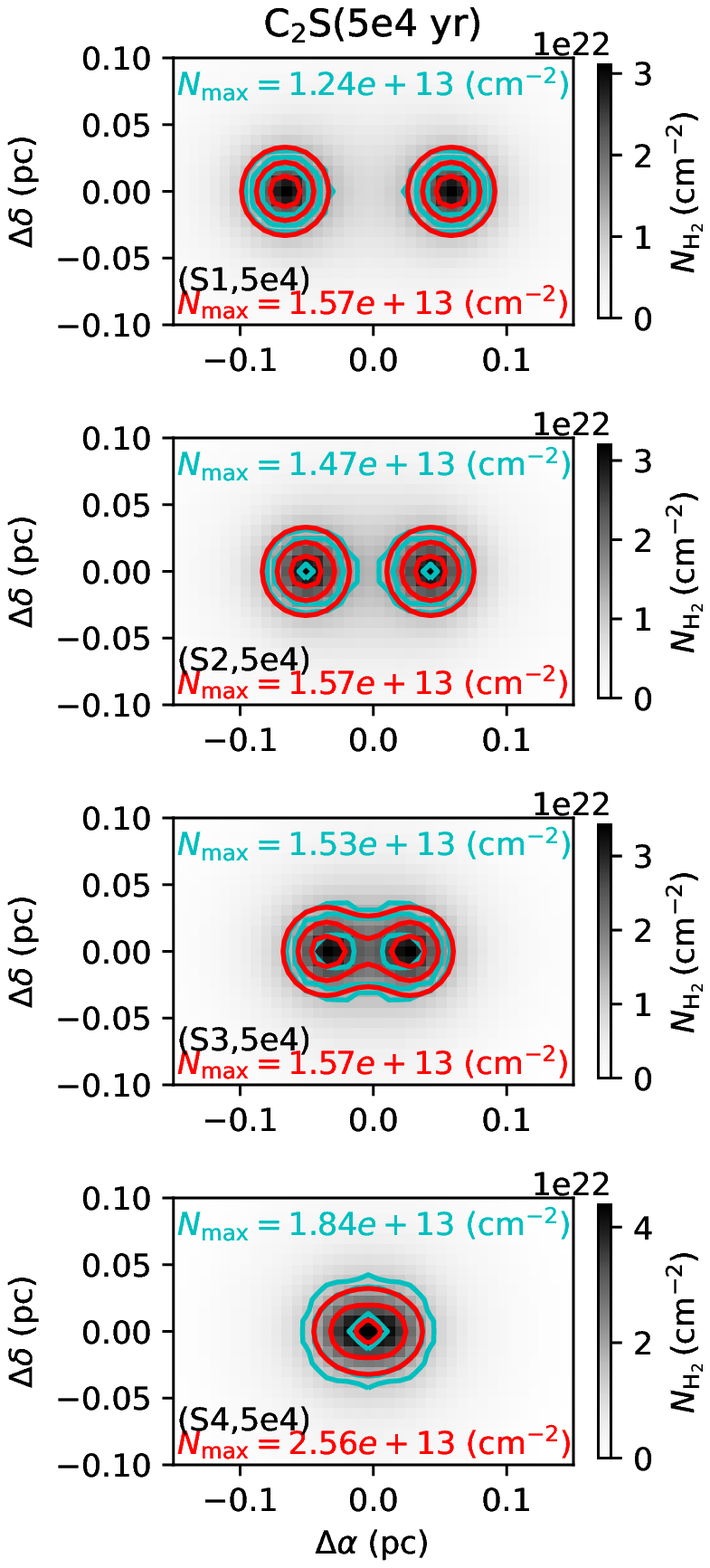}
\includegraphics[width=0.3\textwidth]{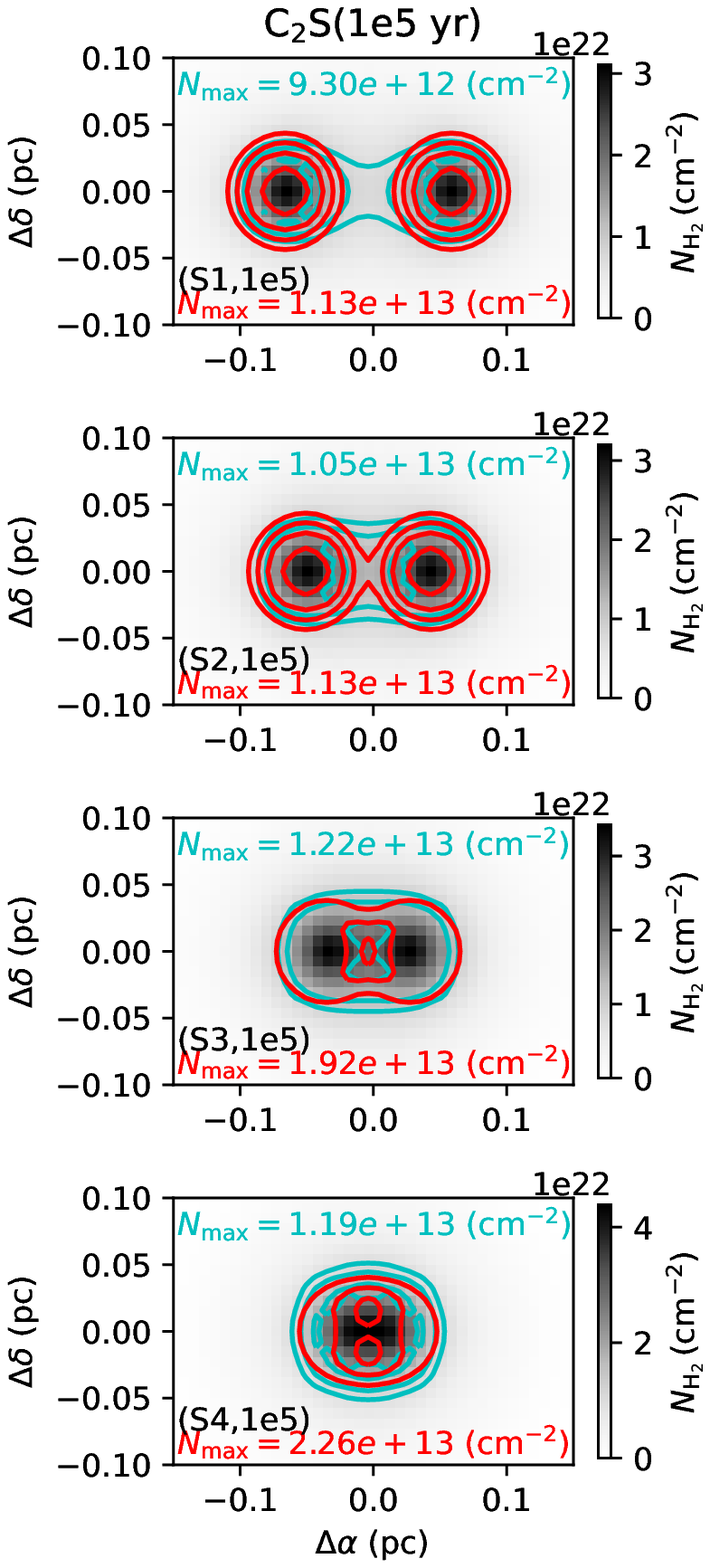}
\includegraphics[width=0.3\textwidth]{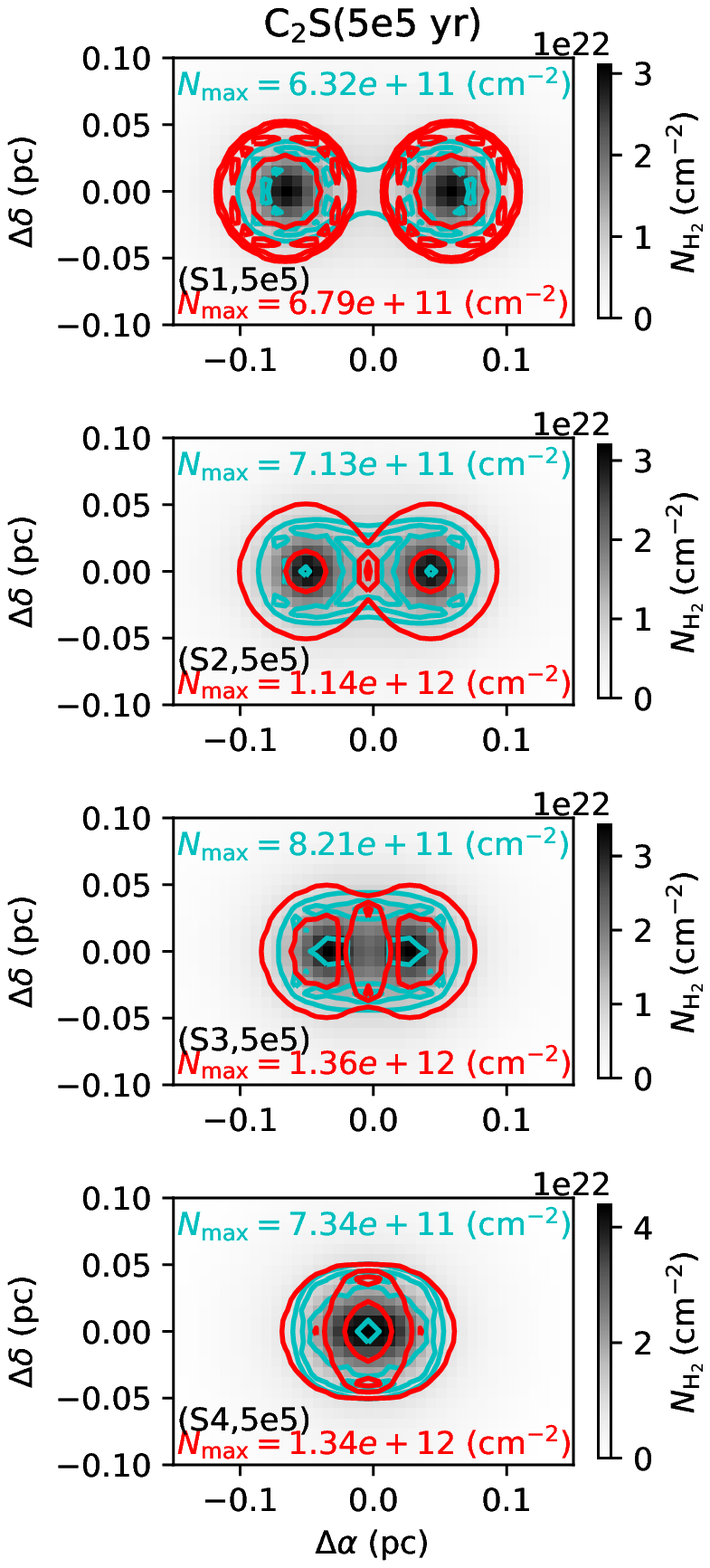}
\caption{Same as Fig.~\ref{fig:N2H+}, but for \ce{C2S}.}
\label{fig:C2S}
\end{figure}

\begin{figure}[ht!]
\centering
\includegraphics[width=0.3\textwidth]{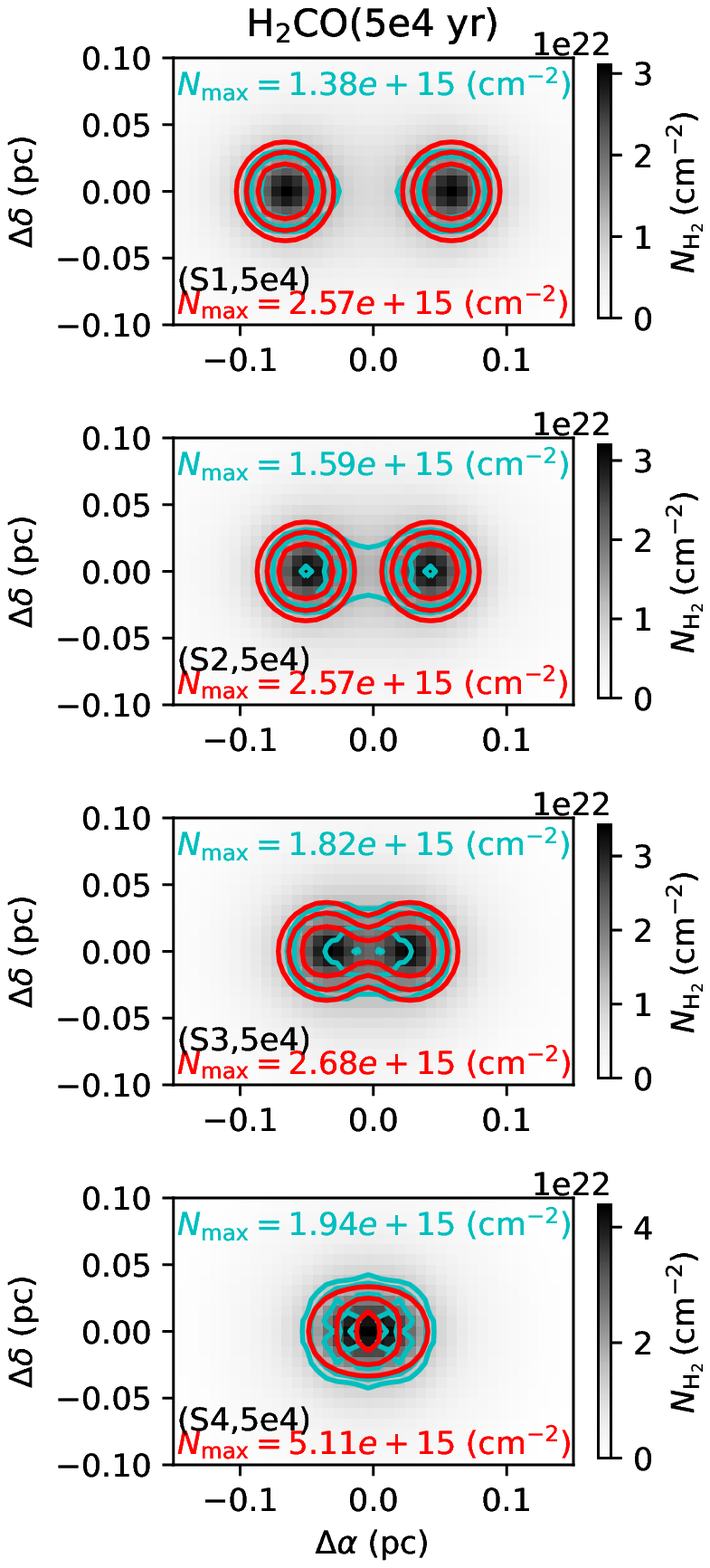}
\includegraphics[width=0.3\textwidth]{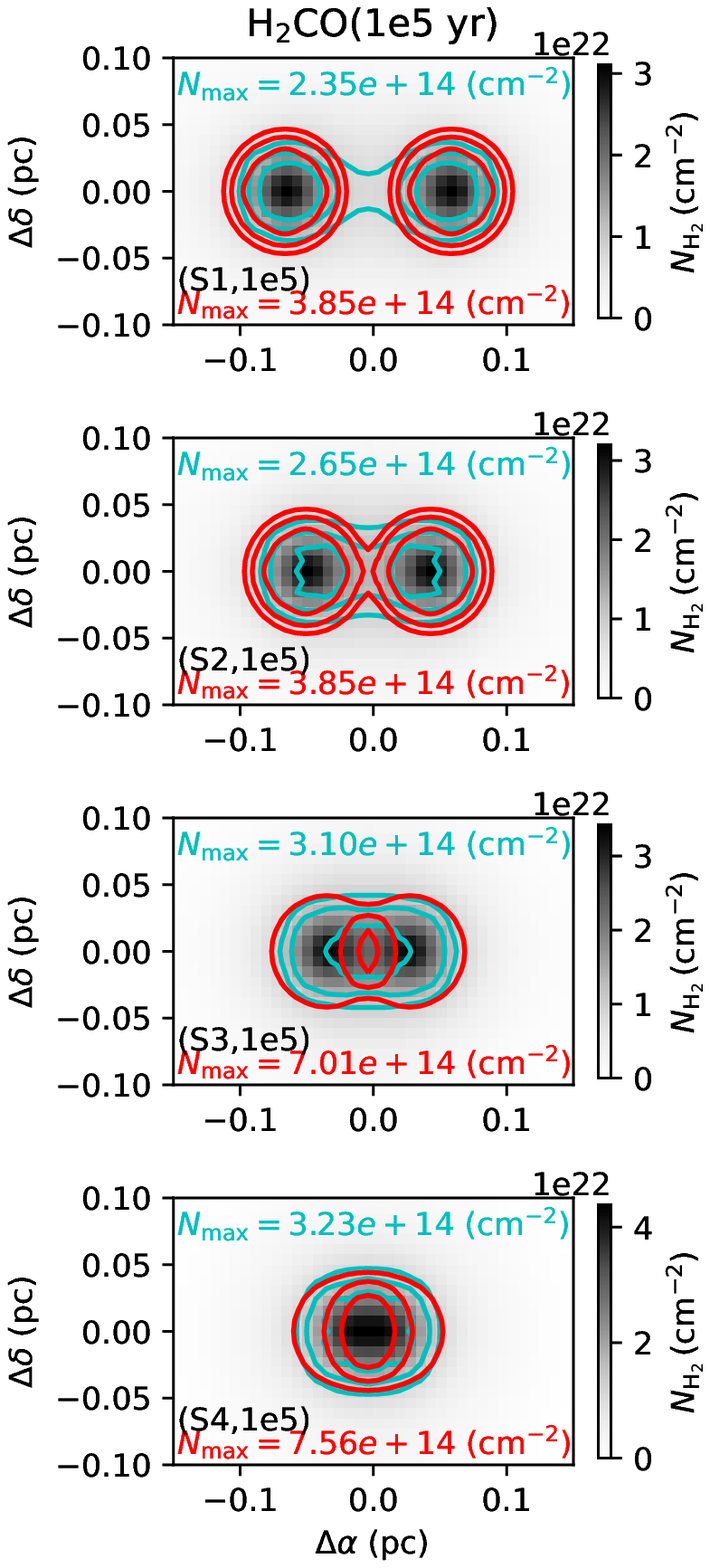}
\includegraphics[width=0.3\textwidth]{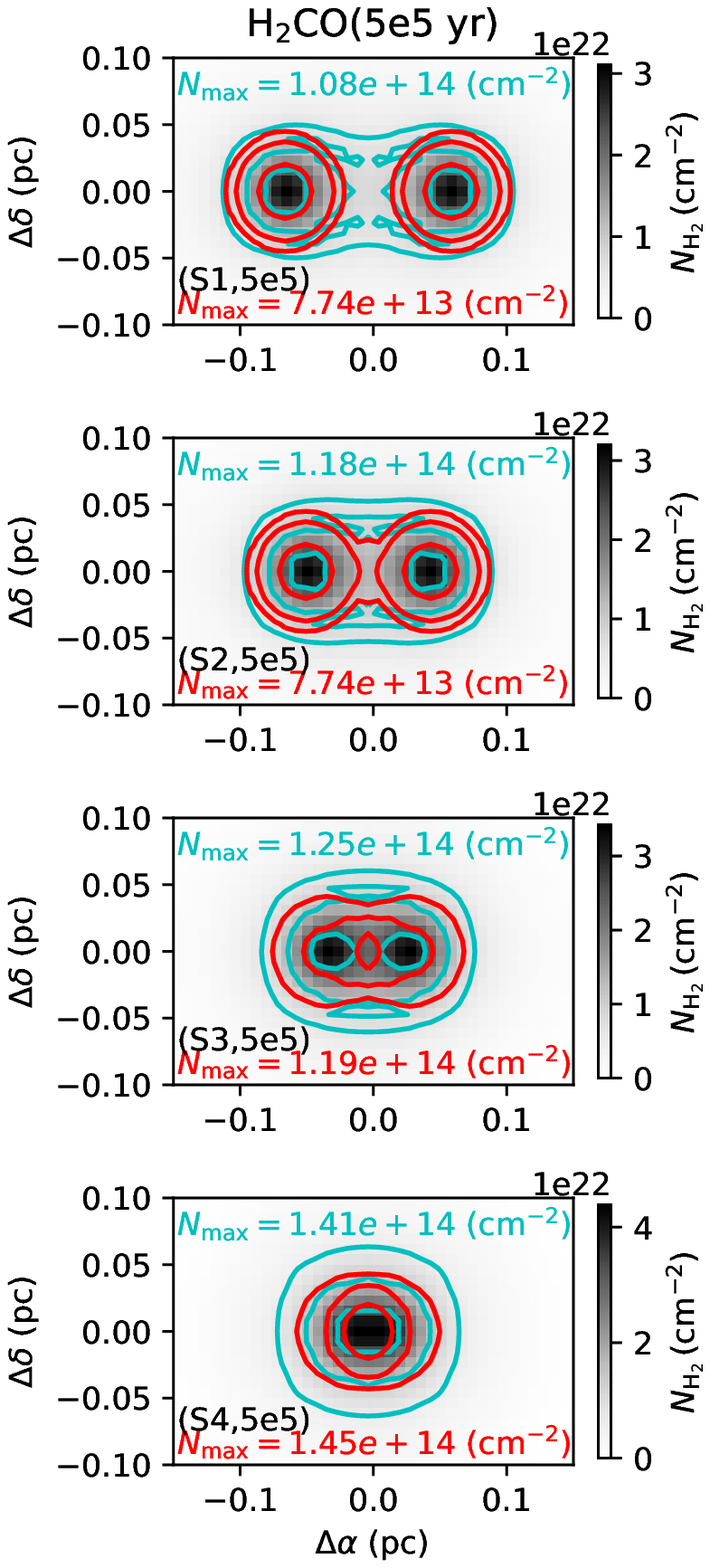}
\caption{Same as Fig.~\ref{fig:N2H+}, but for \ce{H2CO}.}
\label{fig:H2CO}
\end{figure}

\begin{figure}[ht!]
\centering
\includegraphics[width=0.3\textwidth]{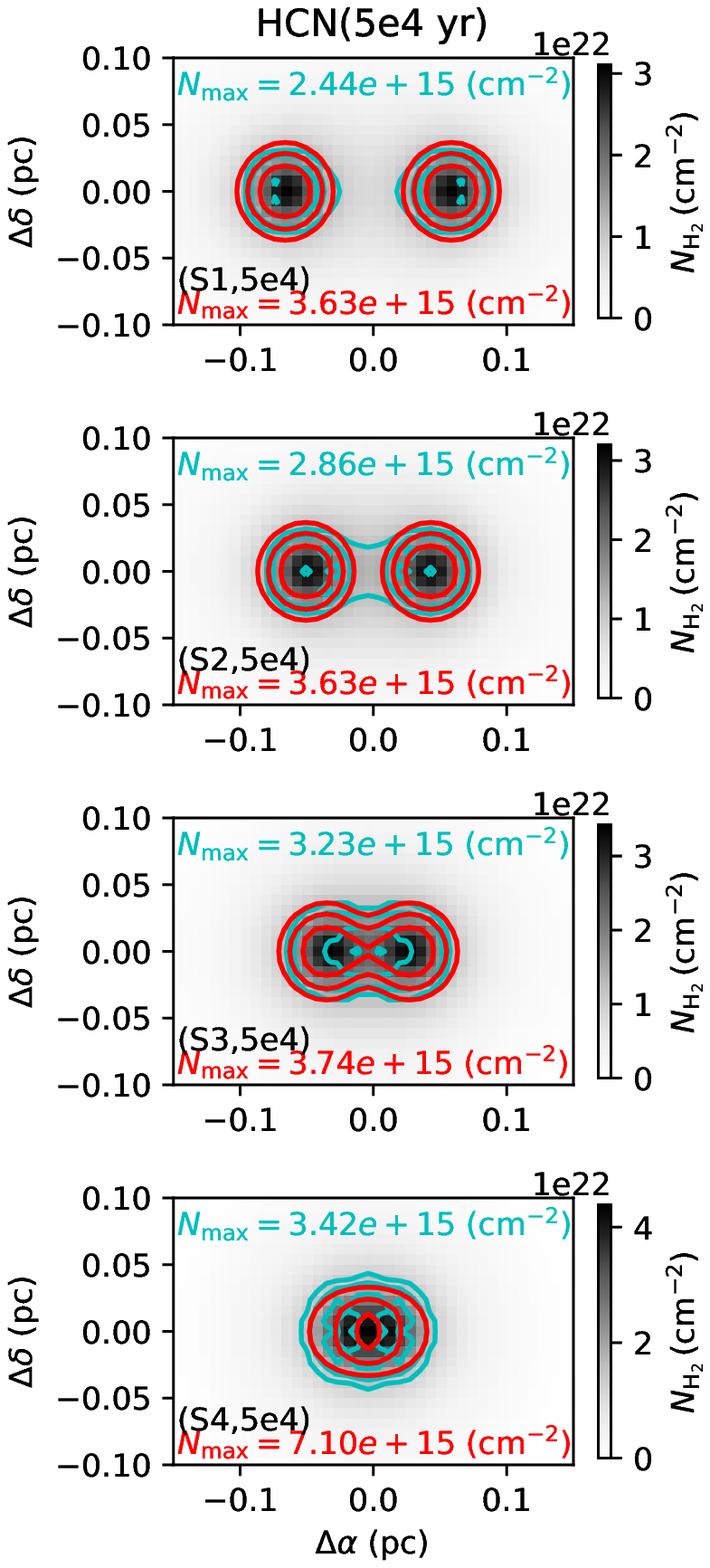}
\includegraphics[width=0.3\textwidth]{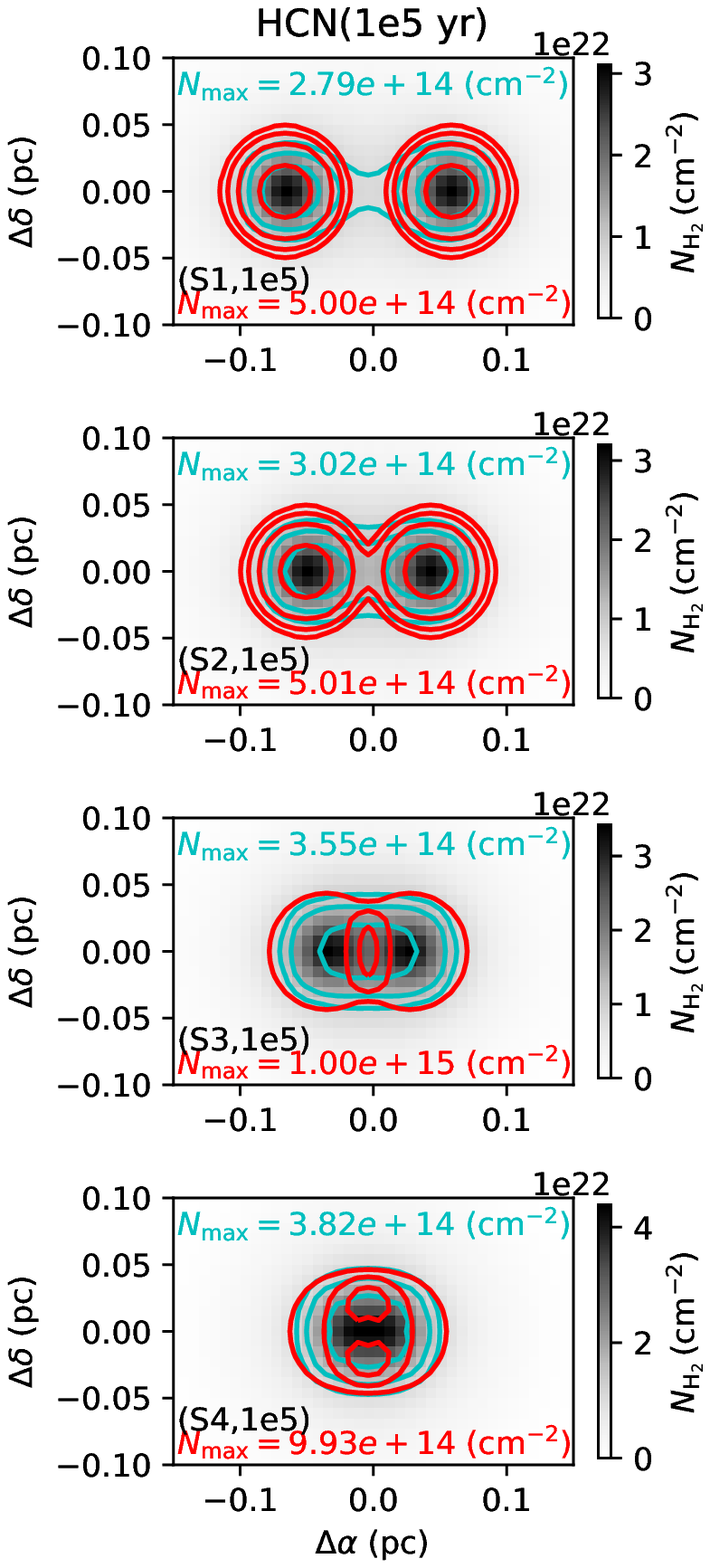}
\includegraphics[width=0.3\textwidth]{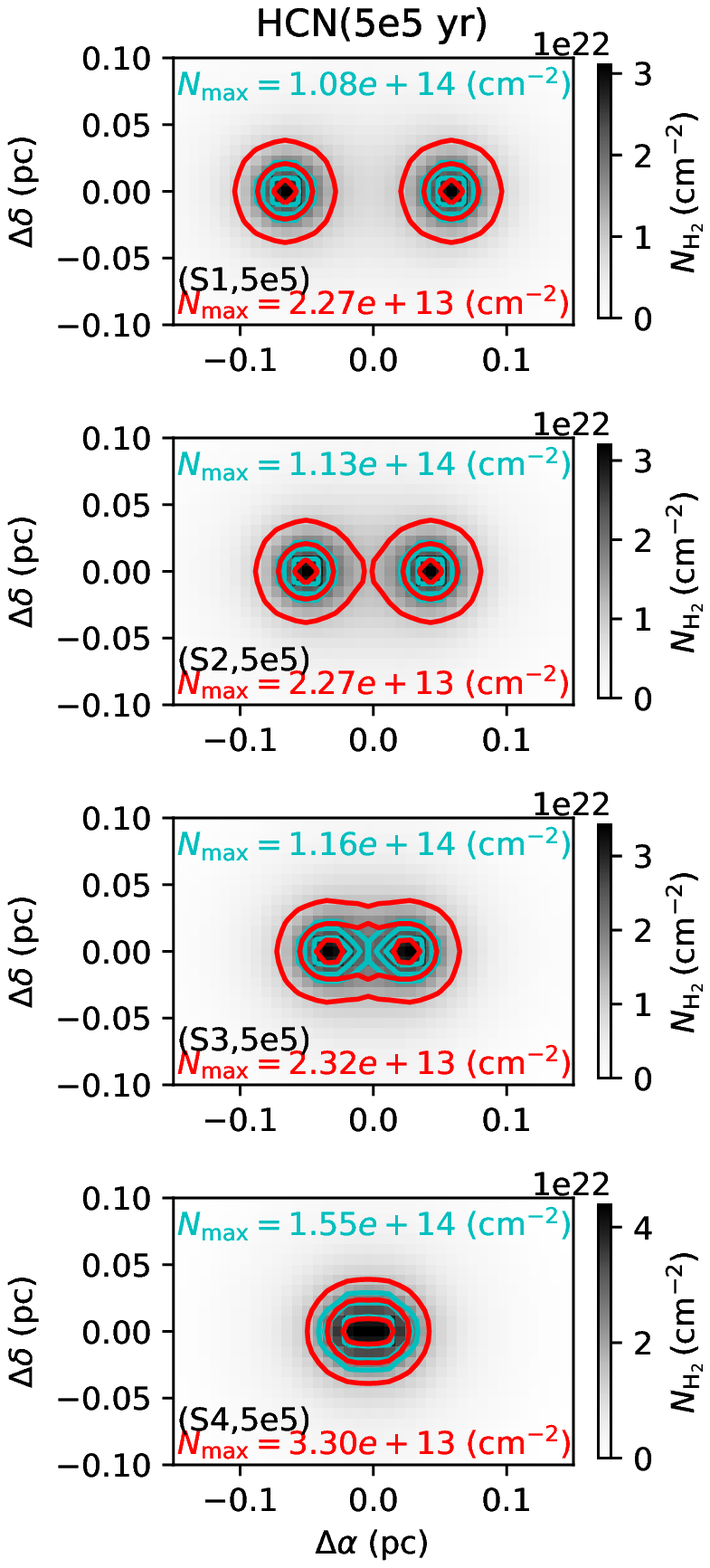}
\caption{Same as Fig.~\ref{fig:N2H+}, but for \ce{HCN}.}
\label{fig:HCN}
\end{figure}

\begin{figure}[ht!]
\centering
\includegraphics[width=0.3\textwidth]{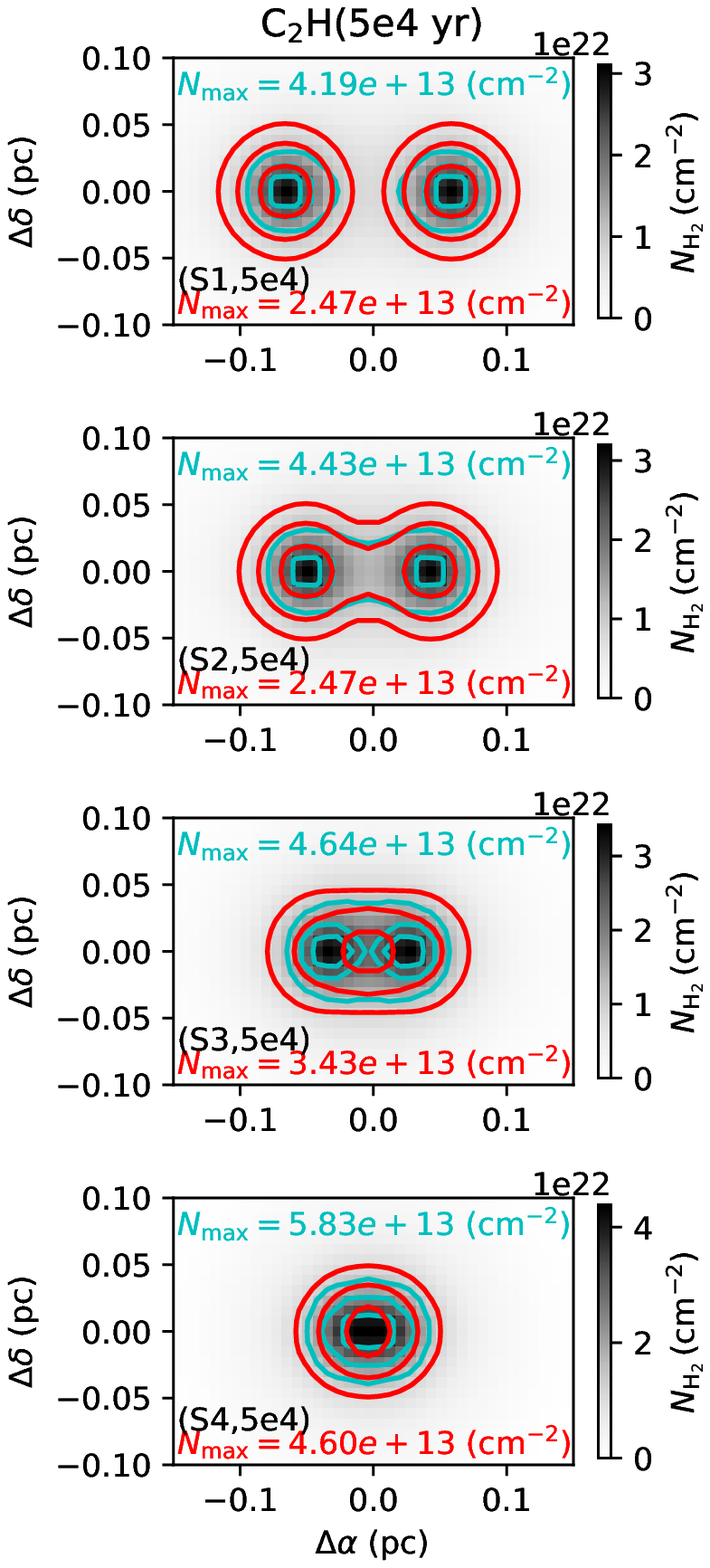}
\includegraphics[width=0.3\textwidth]{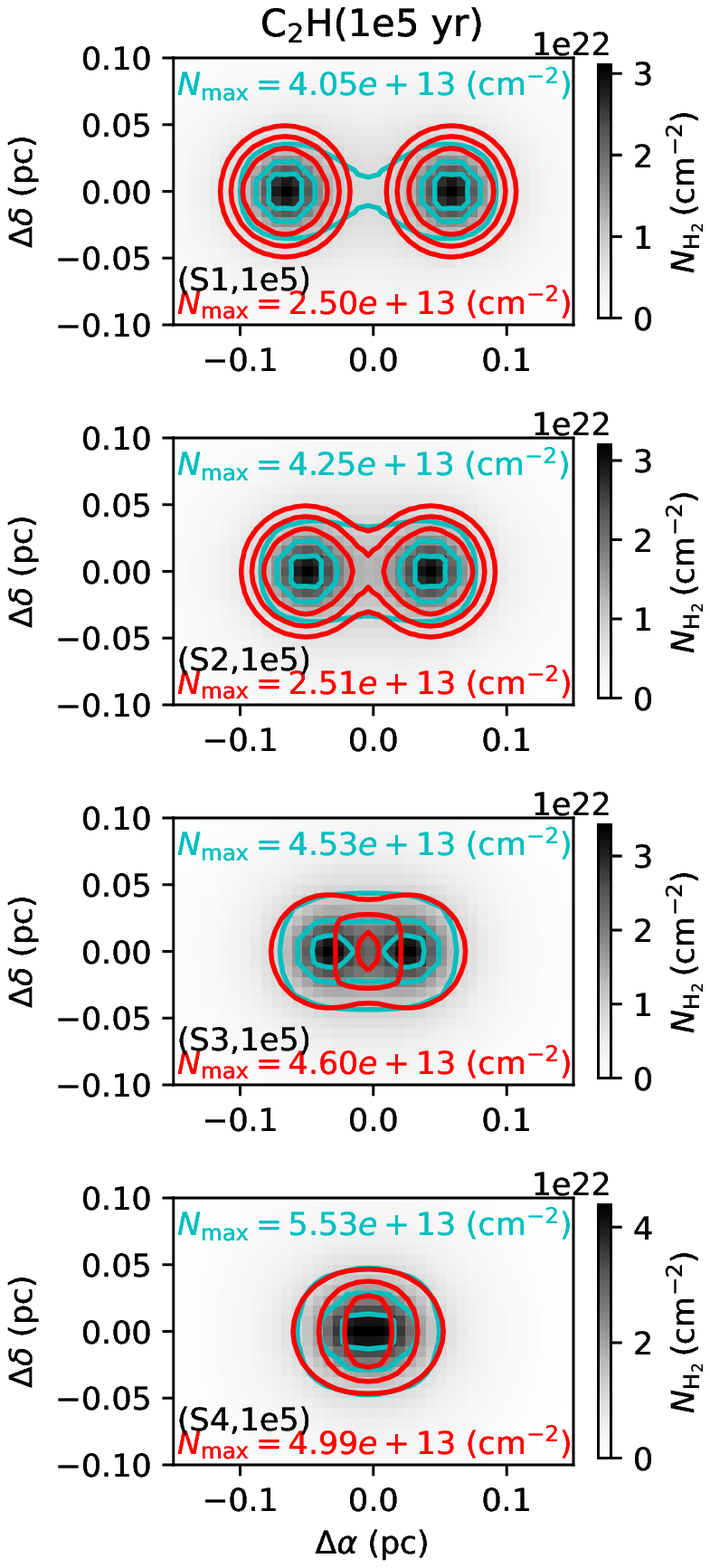}
\includegraphics[width=0.3\textwidth]{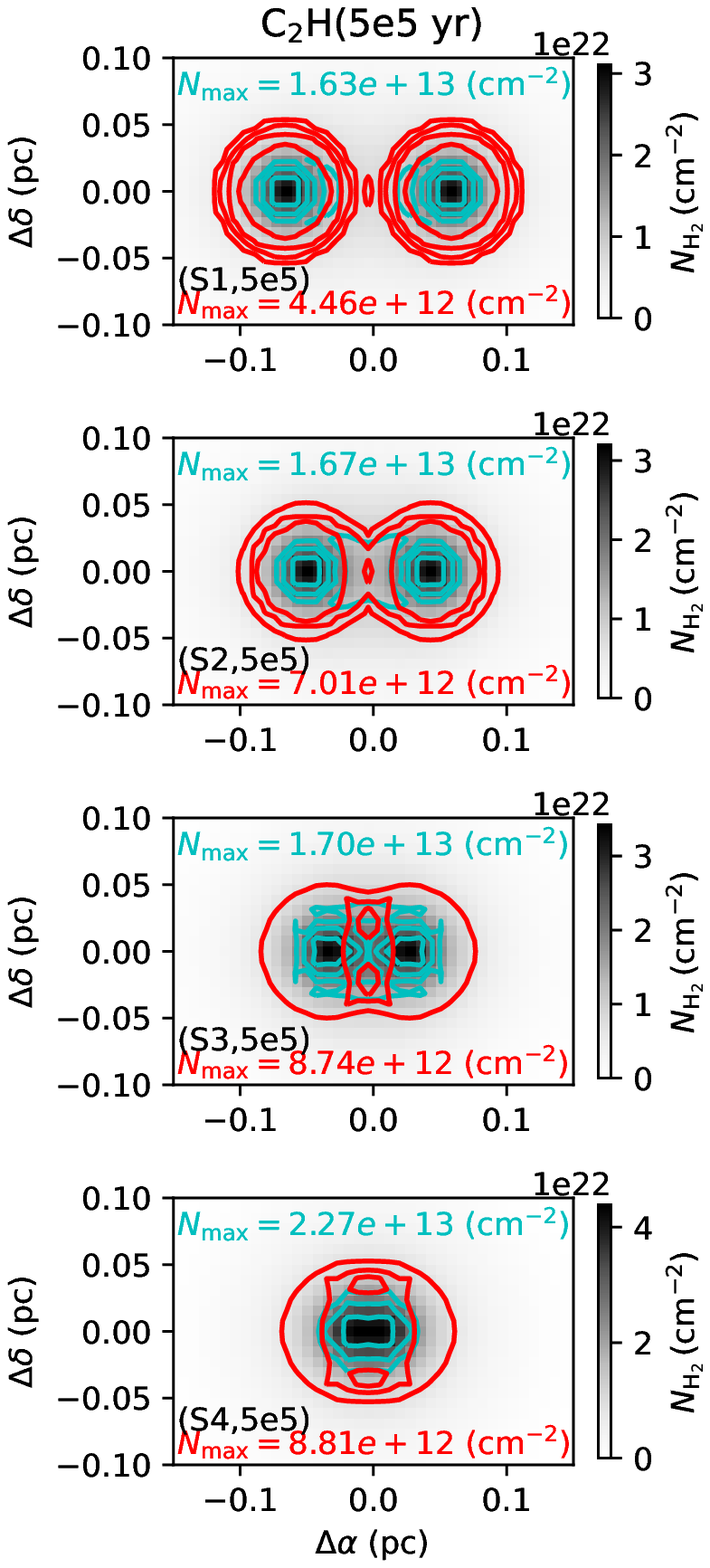}
\caption{Same as Fig.~\ref{fig:N2H+}, but for \ce{C2H}.}
\label{fig:C2H}
\end{figure}

For \ce{HC3N} in Fig.~\ref{fig:HC3N} and \ce{C2S} in Fig.~\ref{fig:C2S}, two big differences between TMCC(red) and TOCC (red) models are found at later ages of $10^5$ and $5\times 10^5$\,yrs when the depletion effects take effect. (1) At age of $10^5$\,yr, the depletion effects start to occur resulting in a small ring structure around each core in the TOCC model. Thus, there are two peaks (red in panels of (S4,1e5)) with linked axis perpendicular to the x-axis; (2) At a later age of $5\times 10^5$\,yr, the peak offset from the two \ce{H2} peaks in the TOCC model (red in panel of (S3,5e5)) is also due to the projection effects. For the TMCC model, depletion effects are larger at the merged parts due to the higher densities, which results in no molecular peak here. Similar behaviors are also found for \ce{H2CO}, HCN and \ce{C2H} shown in Fig.~\ref{fig:H2CO}, \ref{fig:HCN} and \ref{fig:C2H} respectively. These results are also in accord with the modeling of PGCC in paper of \cite{Ge+etal+2020b}, though the physical models are different.

Molecular peak offsets can also be observed at the middle of the two cores in the TMCC model at early ages, such as that (cyan) shown in panel (S3,5e4) of Fig.~\ref{fig:HC3N}, \ref{fig:H2CO} and \ref{fig:HCN} for \ce{HC3N}, \ce{H2CO} and HCN respectively. The molecular peaks are due to that, at the peak position, the depletion does not occur at so early ages. But for the continuum peaks (gray) with higher density, the depletion starts to take effect.
  
\begin{figure}[ht!]
\centering
\includegraphics[width=0.4\textwidth]{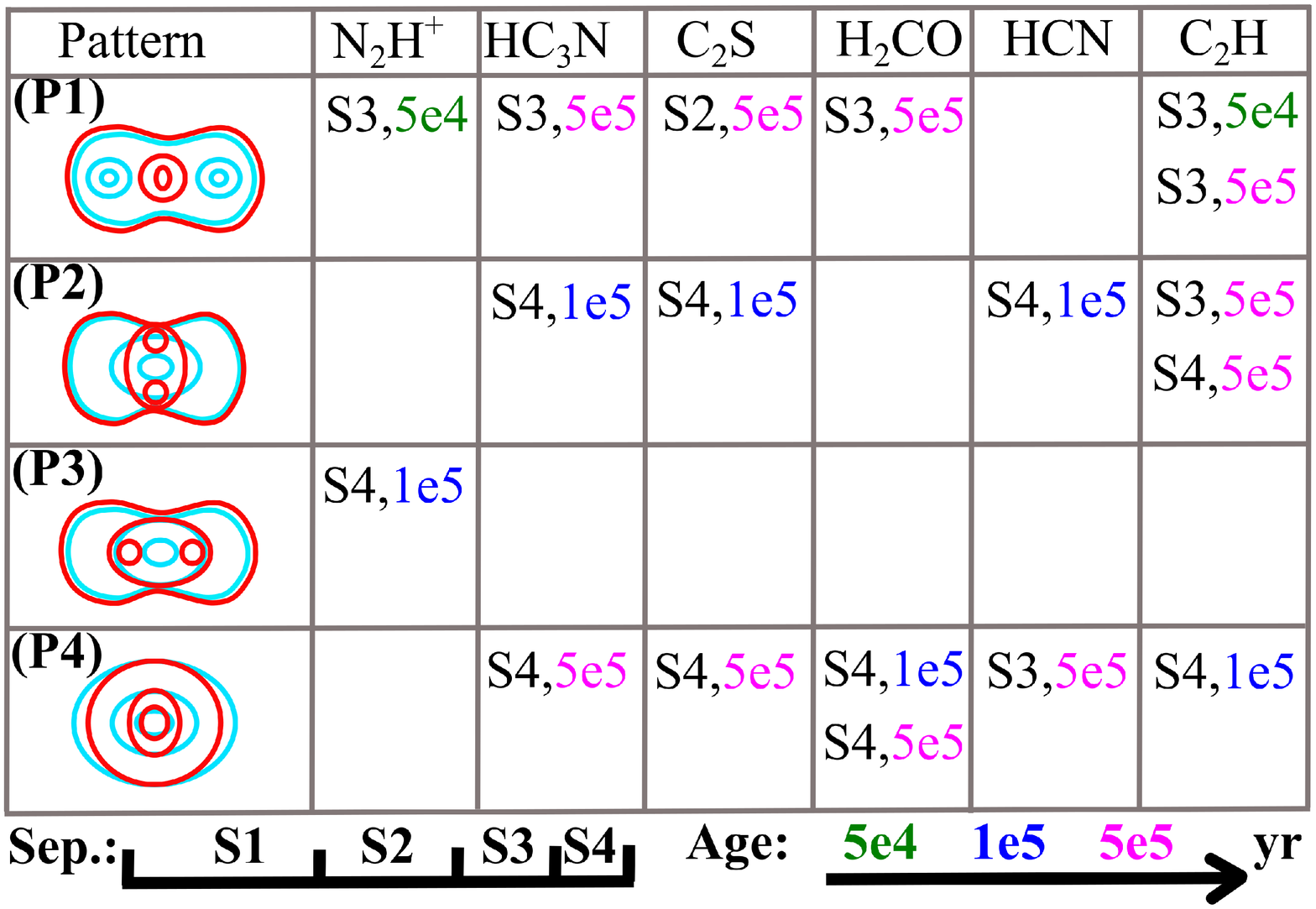}
\caption{Summary of four typical patterns (P1/2/3/4) of molecular distribution differences between the two overlapping cloud core model (red) and the two merging cloud core model (cyan). The separations (S1=0.115, S2=0.085, S3=0.054 and S4=0.023\,pc) and chemical ages ($t=$5e4, 1e5 and 5e5\,yr) of corresponding species that shown in Fig.~\ref{fig:N2H+}-\ref{fig:C2H} are summarized in the right part.}
\label{fig:differences}
\end{figure}

Finally, we summarize typical molecular differences between the TOCC and TMCC models in the left-most part of Fig.~\ref{fig:differences} which shows four patterns marked by names of P1, P2, P3, and P4. In the right parts of Fig.~\ref{fig:differences}, to clearly show the differences as function of separation and chemical age, the panel's ID in Figs.~\ref{fig:N2H+}-\ref{fig:C2H} (e.g. (S1,1e5)) are listed for the above species accordingly. By checking Fig.~\ref{fig:differences}, we find that, for \ce{N2H+}, the patterns P1 and P3 are observed at early ages of 5e4 and 1e5\,yrs. However for \ce{HC3N}, \ce{C2S}, \ce{H2CO}, HCN and \ce{C2H}, patterns P1, P2 and P4 are observed at later ages of 1e5 and 5e5\,yrs. For all the species except for \ce{N2H+}, the patterns are usually observed for the two cores with separations of S3 (0.054\,pc) and S4 (0.023\,pc). The above trends indicate that the peaks observed in the TOCC models strongly depend on the ring size of molecules, e.g the radial peak of \ce{C2S} at $\sim 0.05$\,pc at a later age of about $1\sim 5\times 10^5$\,yr (see e.g. Fig.~\ref{fig:temperature}). This also means that true physical structures from observations are the key points to reproduce observed molecular distributions by applying the projection effects.

In summary, the molecular peak offsets from the \ce{H2} map peaks can be observed at many evolutionary stages (ages) due to the projection effects of the two cores in 3-D space which show different distributions as that of two merging cloud cores. Therefore, the molecular peak offsets from continuum peaks can be used to verify the TDPEs and then to explore 3-D structures. Applied to PGCCs, the peak offsets can be up to $\sim 0.05$\,pc corresponding to $\sim 24.5''$ with the distance of PGCCs in Orion A and B (420\,pc) which are observable. 

\subsubsection{Application: a map with randomly distributed multiple cloud cores}
\begin{figure}[ht!]
\centering
\includegraphics[width=\textwidth]{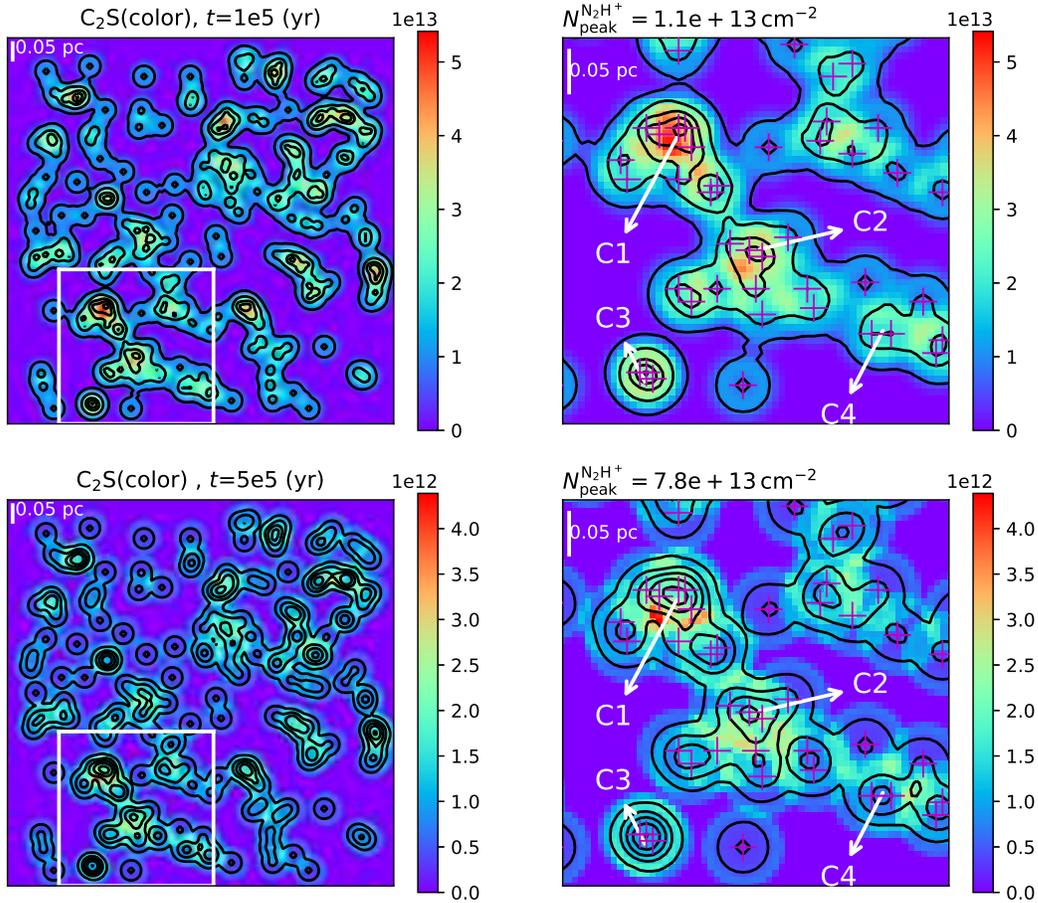}
\caption{\ce{C2S} column density map (color) with projection effects of random distributions of 200 cores at ages of $10^5$ (upper panels) and $5\times 10^5$\,yrs (lower panels). Right panels show enlarged regions in the white boxes in the left panels. The black contours show the \ce{N2H+} map with levels of [0.1, 0.3, 0.5, 0.7, 0.9]$\times N_{\rm peak}^{\rm N_2H^+}$. The values of $N_{\rm peak}^{\rm N_2H^+}$ are labeled in the titles of the right panels. The purple plus signs indicate the core centers before the projection process. A white scale bar is plotted in the upper-left corner of the panels. Four cores (C1/2/3/4) are marked for discussions in the text.}
\label{fig:map:C2S}
\end{figure}
To show the projection effects of molecules with multiple cloud cores, we present a synth map with 200 random cores in a $1.5\times 1.5$\,pc$^2$ region. The spherical cloud core described in Section~\ref{sec:SCC} is used to build the 200 cores in the region. Then the projection process described in Section~\ref{sec:TOCC} is used to make the synth map using random positions of the 200 cores in the region.

We choose \ce{C2S} as our main example because that (1) at an early age of $10^5$\,yr, the depletion effect is small at each core center which can be used as a non-depletion example; (2) At a later age of $5\times 10^5$\,yr, the depletion is strong resulting in a ring structure of \ce{C2S} which is a typical case. For comparison, a continuum map is indicated using \ce{N2H+} map from the same model, which is a well known dense core tracer that can be used to roughly indicate the continuum peaks. In addition, \ce{N2H+} is roughly optically thin which allows the projection process to make the map. We also take \ce{HC3N} as another example to briefly compare with \ce{C2S} and previous observations below. 

Fig.~\ref{fig:map:C2S} shows the synth map with \ce{C2S} as color and \ce{N2H+} as black contours. As have mentioned above, two ages of $10^5$ and $5\times 10^5$\,yrs are shown in the upper and lower panels respectively. The left and right panels show the whole map and an enlarged region to show typical distributions, respectively. From the right-upper panel, we see that \ce{C2S} and \ce{N2H+} have similar peak positions (e.g. at positions of cores C1 to C4) at the early age of $10^5$\,yr at which the depletion of \ce{C2S} is not strong in each core. However from the right-lower panel at the later age of $10^5$\,yr, \ce{C2S} peak offsets from the \ce{N2H+} peaks are obvious (e.g. at positions of cores C1 to C4, and the cores between C1 and C2, between C2 and C4, and between C2 and C3) which are due to the strong depletion of \ce{C2S} in each core. The clumpy structures of \ce{C2S} are also notable which are usually observed in real interstellar clouds such as, observed \ce{C2S} in a sample of PGCCs (\citealt{Tatematsu+2017}).

\begin{figure}[ht!]
\centering
\includegraphics[width=\textwidth]{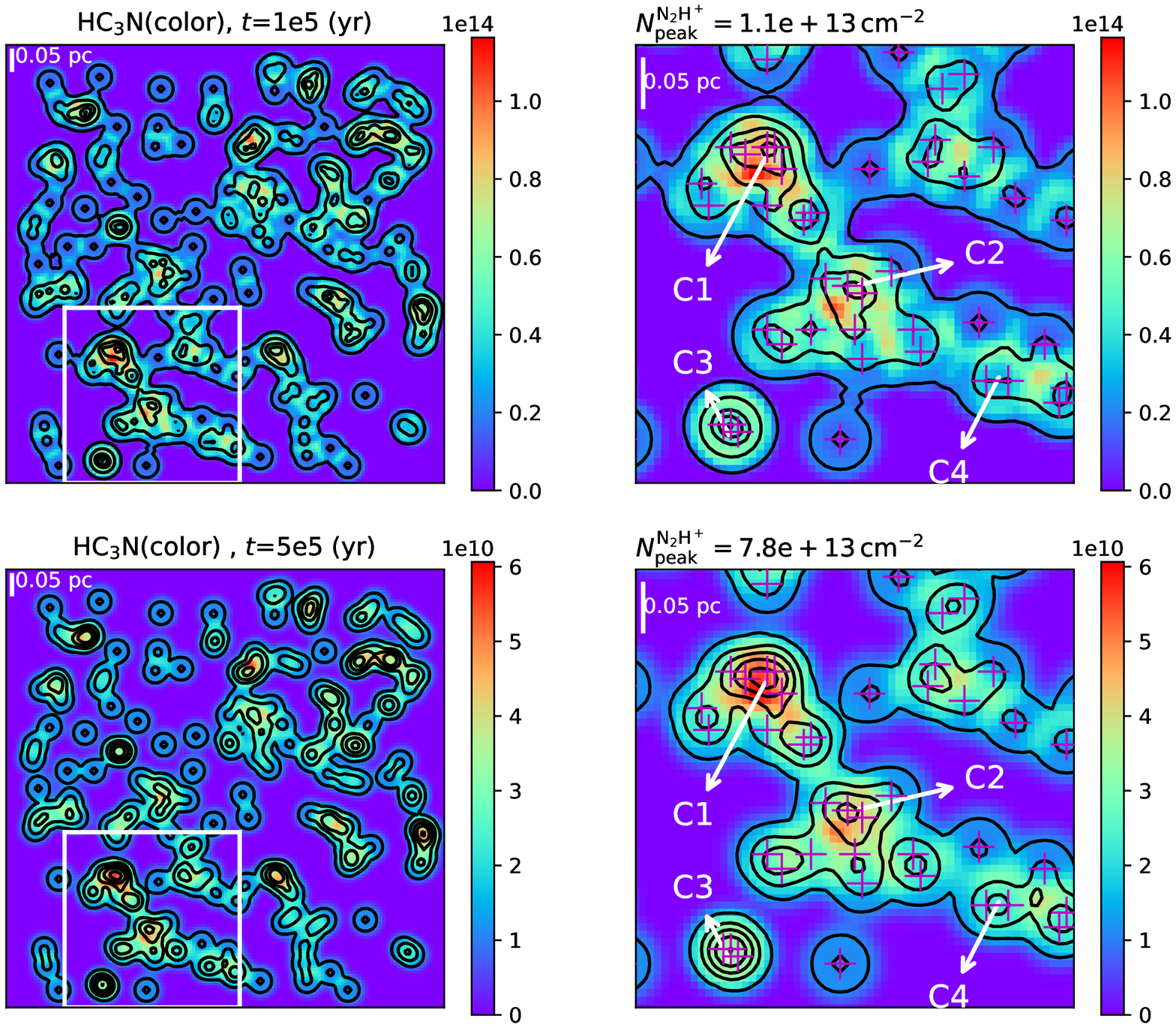}
\caption{Same as Fig.~\ref{fig:map:C2S} but with color scale for \ce{HC3N}.}
\label{fig:map:HC3N}
\end{figure}

The \ce{HC3N} map is shown with color scale in Fig.~\ref{fig:map:HC3N} which shows that \ce{HC3N} has similar peak positions as that of \ce{N2H+} (contours) at the later age ($5\times 10^5$\,yr, lower panels) due to its weaker depletion effect than that of \ce{C2S}. Many other molecules also show similar distributions as that of \ce{C2S} and/or \ce{HC3N} such as \ce{H2CO}, \ce{HCN}, \ce{C2H} and \ce{SO}.% shown in Fig~\ref{fig:map:others}.

Typically, the modeled map show similar distributions to observed maps of \ce{N2H+}, \ce{HC3N} and \ce{C2S} in some observations: 
\begin{itemize}
\item At the early age, our modeled \ce{C2S} and \ce{N2H+} have similar peak positions (upper panels in Fig.~\ref{fig:map:C2S}) as that in TUKH021 in Orion A cloud shown in Fig.3 of \cite{Tatematsu+etal+2010} (hereafter T10), showing similar peak positions of the two molecules. At the later age, our modeled \ce{C2S} show peak offsets from \ce{N2H+} peaks (lower panels Fig.~\ref{fig:map:C2S}) as that observed in TUKH003 in Fig.2 and TUKH122 in Fig.7 of T10, showing \ce{C2S} peak offsets from \ce{N2H+} peaks. Similar distributions were also observed in samples of PGCCs (e.g. \citealt{Tatematsu+2017,Tatematsu+etal+2021}).
\item At the two ages, our modeled \ce{C2S} and \ce{HC3N} (see color scales in Fig.~\ref{fig:map:C2S} and Fig.~\ref{fig:map:HC3N}) have similar distributions (see distributions toward core C1) to that observed in dark cloud L1147 by \cite{Suzuki+etal+2014} (hereafter S14), see their Fig.2 and 3, showing that \ce{HC3N} has similar peak position as the dust continuum map while \ce{C2S} not. The age range in our model is also consistent with the proposed age of $\sim 10^5$\,yr by the model of S14.
\end{itemize}

With the synth map of 200 cloud cores, we have shown that the projection effects could produce similar molecular distributions as that of observations. Thus, it has great potential to explain real observations and to explore 3-D structures through molecular distributions. We also note that the synth map is very rough since we assumed that all the 200 cores are independent objects in 3-D space which allows the projection process. In real interstellar clouds, complex 3-D structures could exist which means that merging cores and overlapping cores could co-exist in the same region. Another point is that there are multiple cores in a dense region (see purple plus signs in Fig.~\ref{fig:map:C2S} towards to the C1 core) which means that complex 3-D structure in the line-of-sight exits but maybe missed by observations. Finally, we used the same physical structure for all cores which maybe not be reasonable in real space. All of these need more high-resolution observations of more molecular lines to constrain the models and to explore the true 3-D structures.

\subsection{Discussions}\label{sec:discussions}
\subsubsection{Temperature effects}\label{sec:temperature}
\begin{figure}[ht!]
\centering
\includegraphics[width=0.5\textwidth]{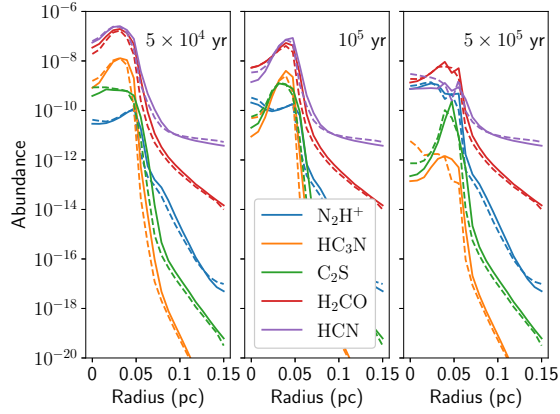}
\caption{Temperature effects on the radial abundance profiles of \ce{N2H+}, \ce{HC3N}, \ce{C2S}, \ce{H2CO} and \ce{HCN}, at ages of $5\times 10^4$ (left), $10^5$ (middle) and $5\times 10^5$\,yrs (right). The solid lines are from the models with varied gas and dust temperatures described in Section~\ref{sec:SCC}. The dashed lines are from the models with a fixed temperature of 10\,K for both gas and dust.}
\label{fig:temperature}
\end{figure}
In this subsection, we demonstrate that the above 3D projection effects are not affected by the use of different gas and dust temperatures in TOCC and TMCC models. We recall that radial profiles of dust and gas temperatures computed in Sect.~\ref{sec:SCC} are used in the TOCC models while homogeneous gas and dust temperatures (10\,K) are adopted for the TMCC models. To examine the potential impact of the different temperatures on our discussions, we set up a new TOCC model by replacing the radial gas and dust temperature profiles with the same homogeneous temperature of 10\,K as in the TMCC models. Fig.~\ref{fig:temperature} shows the radial abundances of \ce{N2H+} \ce{HC3N}, \ce{C2S}, \ce{H2CO} and \ce{HCN}, at ages of $5\times 10^4$ (left), $10^5$ (middle) and $5\times 10^5$\,yrs (right). The solid and dashed lines are from models with varied temperature and fixed temperature respectively. 

From Fig.~\ref{fig:temperature}, we see that most species at all the selected ages have very small differences between solid and dashed lines which means that the temperatures used in our TOCC and TMCC models do not change our above conclusions. However, for \ce{HC3N} at a later age of $5\times 10^5$\,yr (yellow in the right panel), a higher temperature of 10\,K (dashed) enhances its abundance due to that it is linked to \ce{N2} which has low desorption temperature from dust grains and high gas-phase abundance (\citealt{Womack+etal+1992}). The \ce{N2H+} and \ce{HCN} are also tightly linked to \ce{N2}, but they have bigger gas-phase abundances than that of \ce{HC3N}. This is the reason why only \ce{HC3N} have big enhancements at 10\,K. Since the temperature variations ($10-15$\,K) do not affect the modeled radial peaks of molecules (e.g. \ce{N2H+}, \ce{HC3N} and \ce{C2S}) in chemical models as shown in Fig. 10 of \cite{Ge+etal+2020a}, we do not test these temperatures in this work. In summary, the temperatures used in our TOCC and TMCC models do not affect our conclusions on the molecular distribution differences.

\subsubsection{Isotopes}
The tracers of our SPACE project also include isotopes (e.g. \ce{C^{18}O} and \ce{DNC}), which is not included in the current models. However, we can expect similar distributions of isotopes as that of their main forms because that D-bearing and \ce{^{13}C}-bearing species have similar evolutionary trends (depletion effects) (see e.g. modeled \ce{NH3} and \ce{NH2D} in Fig.2, and HNC and \ce{HN^{13}C} in Fig.7 of \cite{Roueff+etal+2015}) and distributions (see e.g. observed CO and \ce{^{13}CO} in a sample of PGCCs by \cite{Wu+2012}, and \ce{C3H2} and \ce{^{13}CH2} in starless core L1544 by \cite{Spezzano+etal+2017}) to their main isotopes, similar three-dimensional projection effects are also expected.

\subsubsection{Application to real clouds}
In this work, the models are taken as showcases to show that GGCHEMPY code is efficient for building 1-D, 2-D and 3-D simulations taking one typical set of physical parameters of PGCCs. Although we have drawn some useful conclusions from our models, we should keep in mind that complex 3-D structures exist in the real space which may result in complex projection effects (see e.g. projection effects from MHD simulations of \cite{Li+2019}). Application of it to more complicated cloud models is one thing. Another point is that real clouds are clumpy and irregular. This can distort the simplistic pictures given in this paper. Thus, modeling  and observations at higher spatial resolutions will be important to unambiguously recognize the 3D projection  effects and to constrain the 3D structures of real clouds.

\section{Summary}\label{sec:summary}
In this paper, we present the new pure Python-based gas-grain chemical code (GGCHEMPY). By using the Numba package, besides the flexible Python syntax, GGCHEMPY reaches a comparable speed to the Fortran-based versions and can be used for chemical simulations efficiently. 

As a showcase, 1-D, 2-D and 3-D physical and chemical models are shown upon typical conditions of Planck galactic cold clumps to study chemical differences of different physical structures. By comparing the modeled molecular peaks with the \ce{H2} peaks in the overlapping two-core cloud model, we find that the molecular peak offsets from the \ce{H2} peaks can be used to trace the projection effects due to the depletion effects and projection effects and to indicate the evolutionary stages. Compared to the models with a merged two-core cloud model, the molecular distribution differences can be used to distinguish the two 3D cloud structures which have great potential to explain real observations. These simulated peak offsets caused by the three-dimensional projection effects support our scientific goal of the SPACE project towards a sample of Planck Galactic cloud clumps. Future works including isotopes and dust size distribution will be done to interpret observations.

\normalem
\begin{acknowledgements}
This work is accomplished with the support from the Chinese Academy of Sciences (CAS) through a  Postdoctoral Fellowship administered by the CAS South America Center for Astronomy (CASSACA) in Santiago, Chile. JG thanks Dr. Jinhua He and Dr. Tie Liu for their constructive suggestions which significantly improved this paper.
\end{acknowledgements}

\appendix

\section{Interpolation on model grid}\label{app:grid}
To reduce computation time, we introduce an interpolating method on a model grid. We build the single-point chemical model grid with varied density, extinction and temperature which results in about 700 models and needs about 3 hours to run (see Table~\ref{tab:grid}). For each density, the extinction values vary within reasonable values. Thus the temperature values are set to reasonable values according to the correlation between extinction and dust temperature (e.g. $T(A_{\rm V}=0.1)\sim 17$\,K and $T(A_{\rm V}=20)\sim 7$\,K, see \citealt{Hocuk+2017}) with considerations of wide possible uncertainties. We assume the same temperature for gas and dust grains in the models.

\begin{figure}[ht!]
\centering
\includegraphics[scale=0.5]{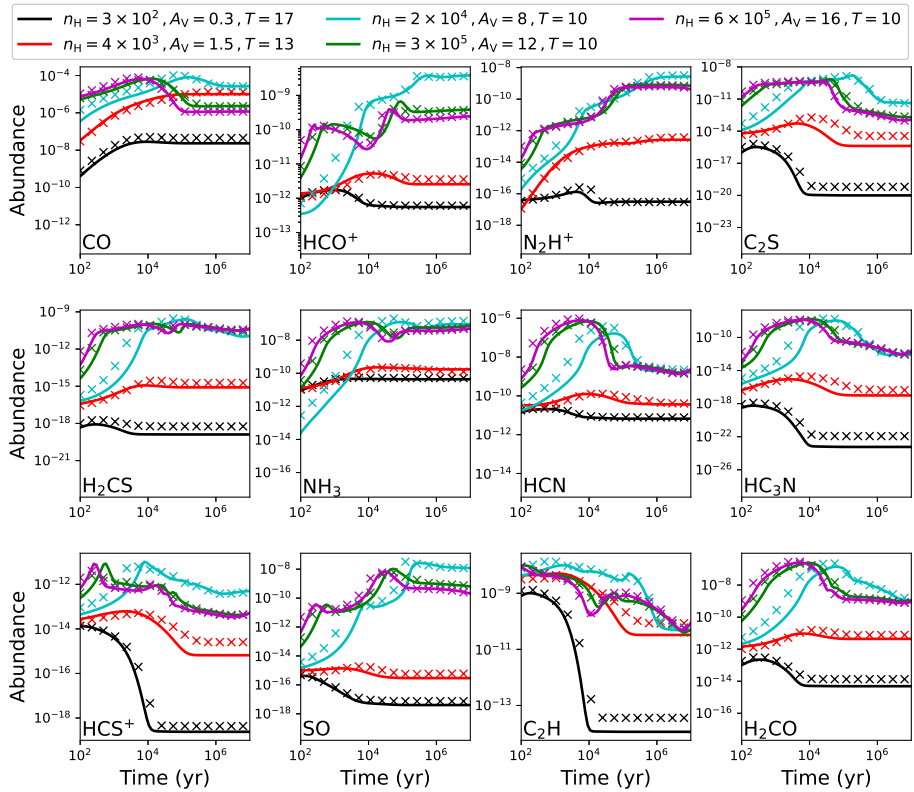}
\caption{Comparison between interpolated values on grids (cross) and the values from true models (line) with five sets of parameters shown in the top legends.}
\label{fig:grid:comparison}
\end{figure}

Therefore, the modeled abundance can be quickly estimated by interpolating on the grid with any given set of density, extinction and temperature. Two strategies are used to do the interpolation of a species at a given age: (1) When the given density is in the grid shown in the first column of Table~\ref{tab:grid}, the species abundance is interpolated on the 2-D extinction-temperature map at the given age; (2) If the given density is not in the density grid, two nearest densities are selected. Thus two abundances values are obtained using the same process used in (1). Finally, the species abundance is the average one of the two abundances. Fig.~\ref{fig:grid:comparison} shows the good agreements between the interpolated values (cross) and the true models (line) which show very small differences.
\begin{table}[ht!]
\centering
\caption{Model grid}
\label{tab:grid}
\begin{tabular}{ccc}
\hline
$n_{\rm H}$\,(cm$^{-3}$) & $A_{\rm V}$(mag) & $T$ (K)  \\
\hline
$10^2$           & 0.1, 0.5, 1, 1.5, 2                        & 18, 17, 16, 15, 14, 13, 12, 11, 10 \\
 $5 \times 10^2$ & 0.1, 0.5, 1, 1.5, 2                        & 18, 17, 16, 15, 14, 13, 12, 11, 10 \\
 $10^3$          & 0.1, 0.5, 1, 2, 3, 4, 5                              & 16, 15, 14, 13, 12, 11, 10 \\
 $5\times 10^3$  & 0.1, 0.5, 1, 2, 3, 4, 5                              & 16, 15, 14, 13, 12, 11, 10 \\
 $10^4$          & 1, 2, 3, 4, 5, 6, 7, 8, 9, 10, 11, 12, 13, 14, 15, 16  & 14, 13, 12, 11, 10, 9, 8\\
 $5\times 10^4$  & 1, 2, 3, 4, 5, 6, 7, 8, 9, 10, 11, 12, 13, 14, 15, 16  & 14, 13, 12, 11, 10, 9, 8\\
 $10^5$          & 5, 6, 7, 8, 9, 10, 11, 12, 13, 14, 15, 16, 17, 18         & 12, 11, 10, 9, 8, 7, 6\\
 $5\times 10^5$  & 5, 6, 7, 8, 9, 10, 11, 12, 13, 14, 15, 16, 17, 18         & 12, 11, 10, 9, 8, 7, 6\\
 $10^6$          & 5, 6, 7, 8, 9, 10, 11, 12, 13, 14, 15, 16, 17, 18         & 12, 11, 10, 9, 8, 7, 6\\
\hline
\end{tabular}
\end{table}

\bibliographystyle{raa}
\bibliography{RAA-2021-0241.R1}

\end{document}